\documentclass[aip,
amsmath,amssymb,
preprint,%
]{revtex4-1}

\usepackage[utf8]{inputenc}
\usepackage[T1]{fontenc}
\usepackage{mathptmx}
\usepackage{dcolumn}

\usepackage{graphicx}
\usepackage{caption}
\usepackage{subcaption}
\usepackage{amsmath}
\usepackage[version=4]{mhchem}
\usepackage{siunitx}
\usepackage{bm}
\usepackage{amssymb}
\usepackage{mathtools}
\usepackage{color}
\usepackage{xspace}
\usepackage{upgreek}
\usepackage{wasysym}
\usepackage{tikz}
\usepackage{xpatch}
\usepackage{verbatim}
\usepackage{mathtools}
\makeatletter
\xpatchcmd{\@float}{\csname fps@#1\endcsname}{h!}{}{}
\makeatother

\newcommand{\norm}[1]{\left\lVert#1\right\rVert}

\newsavebox{\astrutbox}
\sbox{\astrutbox}{\rule[-5pt]{0pt}{20pt}}

\begin{document}

\preprint{AIP/Physics of Fluids}

\title{Role of Pulsatility on Particle Dispersion in Expiratory Flows}
%

\author{K. Monroe}
\affiliation{Department of Aerospace Engineering, University of Michigan, Ann Arbor, MI, 48109}
\author{Y. Yao}
\affiliation{Department of Mechanical Engineering, University of Michigan, Ann Arbor, MI, 48109}
\author{A. Lattanzi}
\affiliation{Department of Mechanical Engineering, University of Michigan, Ann Arbor, MI, 48109}
\author{V. Raghav}
\affiliation{Department of Aerospace Engineering, Auburn University, Auburn, AL, 36849-0001}
\author{J. Capecelatro}
\email{jcaps@umich.edu}
\affiliation{Department of Aerospace Engineering, University of Michigan, Ann Arbor, MI, 48109}
\affiliation{Department of Mechanical Engineering, University of Michigan, Ann Arbor, MI, 48109}

\date{\today}


\begin{abstract}
Expiratory events, such as coughs, are often pulsatile in nature and result in vortical flow structures that transport respiratory particles. In this work, direct numerical simulation (DNS) of turbulent pulsatile jets, coupled with Lagrangian particle tracking of micron-sized droplets, is performed to investigate the role of secondary and tertiary expulsions on particle dispersion and penetration. Fully-developed turbulence obtained from DNS of a turbulent pipe flow is provided at the jet orifice. The volumetric flow rate at the orifice is modulated in time according to a damped sine wave; thereby allowing for control of the number of pulses, duration, and peak amplitude. The resulting vortex structures are analyzed for single-, two-, and three-pulse jets. The evolution of the particle cloud is then compared to existing single-pulse models. Particle dispersion and penetration of the entire cloud is found to be hindered by increased pulsatility. However, the penetration of particles emanating from a secondary or tertiary expulsion are enhanced due to acceleration downstream by vortex structures.
\end{abstract}

\maketitle

\section{Introduction}\label{sec:intro}
\begin{figure}[!h]
\captionsetup[subfigure]{labelformat=simple}
\centering
{\includegraphics[width=.9\textwidth]{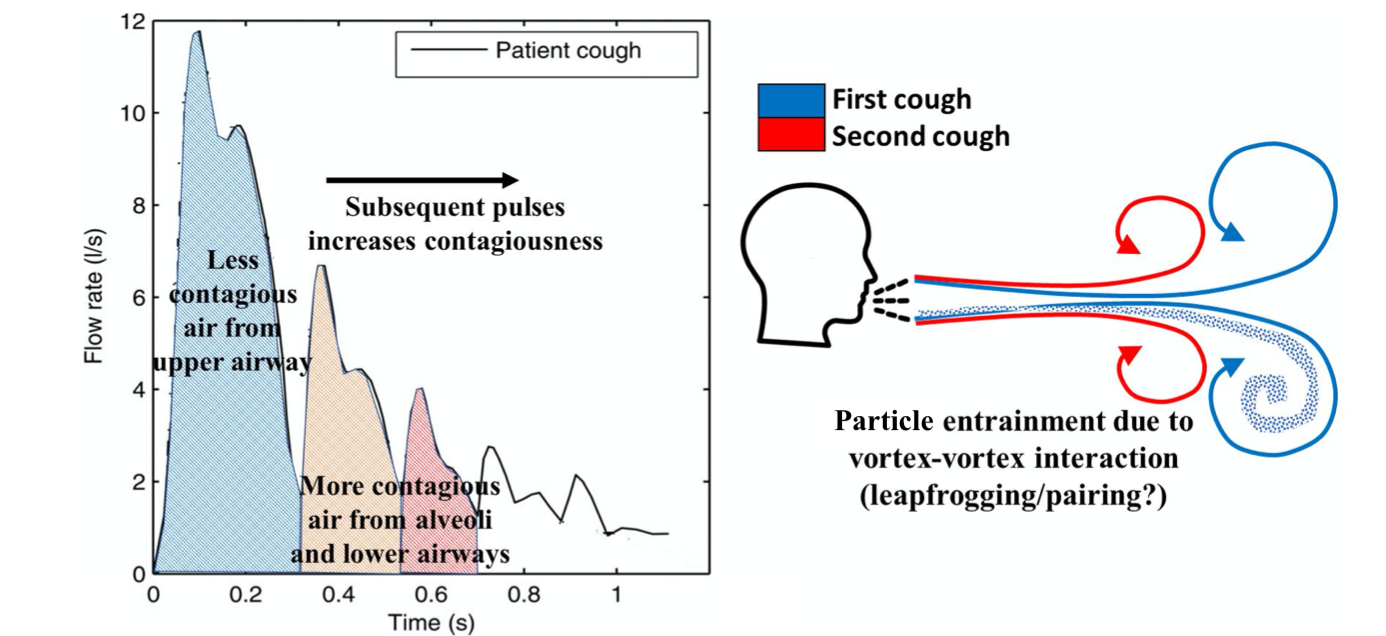}}
\caption{Left: Realistic cough profile from a patient showing expected levels of contagion concentration in the expired air assuming infection is located deeper within the lungs (adapted from \citet{lindsley2013cough}). Right: Hypothesized vortex-vortex interactions leading to increase in particle entrainment.}
\label{fig:motivation}
\end{figure}

Particle transport by turbulent free-shear jets plays a crucial role in many engineering and environmental applications. For example, atomization of liquid fuels leads to complex droplet size distributions and dispersion patterns that strongly influence internal combustion engine efficiency~\cite{aggarwal_review_1998} while pyroclastic density currents, generated by large density differences between gas-particle mixtures, feeds explosive volcanic eruptions~\cite{dufek2016fluid}. Of particular importance during the COVID-19 pandemic is the transmission of liquid droplets and aerosols (referred to interchangeably as particles herein) due to coughing, sneezing, or continuous speech~\cite{mittal_flow_2020}. Accurately describing particle dispersion from expiratory events is a critical aspect to defining physics-informed guidelines for social distancing best practices. While remarkable insight has been gained from analytical~\cite{wells1934air,mittal2020mathematical}, experimental~\cite{bourouiba2014violent}, and computational~\cite{yang_towards_2020,balachandar2020host,aiyer2020coupled} works, the vast majority of studies are restricted to single expulsion events. However, realistic coughing is often characterized by multiple expulsions that lead to vortex-vortex interactions, which can have significant consequences on particle dynamics~\cite{chein1987effects} (see Fig.~\ref{fig:motivation}).

Experimental measurements have demonstrated that realistic coughs are pulsatile, involving a sequence of coughing events, sometimes referred to as ``cough epochs.''~\cite{wei2017penetration,borders2020variability} The flow rate associated with a typical human cough is shown in Fig.~\ref{fig:motivation}. Multiple pulses are observed over a duration of approximately $1$ s, with the peak amplitude occurring at $0.1$ s~\cite{lindsley2013cough}. \citet{gupta2009flow} experimentally characterized the flow dynamics of coughs from human subjects, showing that the flow rate variation of a cough with time can be defined as a combination of gamma probability functions. While single-pulse expiratory events have been well-studied, the influence of pulsatility on both the particle and fluid physics has received significantly less attention.

To understand the influence of pulsatility, it is important to first consider the modes of particle generation and sites of origination~\cite{morawska2009size} during speech, coughing, or sneezing -- Bronchiolar~(droplet size $<1-5\,\upmu$m); Laryngeal~($5-20\,\upmu $m); and Oral~($>50\,\upmu$m)~\cite{wei2016airborne}. If the site of severe infection is deeper in the lungs~(bronchiolar), the expelled aerosols/droplets are generated by a ``fluid-film burst'' mechanism~\cite{johnson2009mechanism} during collapse and reopening of the small airways resulting in small size particles. Since several respiratory infections, including H5N1 and SARS-CoV, replicate primarily in the bronchioles and alevoli~\cite{mossel2008sars,morawska2009size}, the aerosols/droplets generated in the lower airways are likely to contain higher doses of virus particles. In this case, secondary and tertiary pulses of a multi-pulse cough will expel the volume of air originating predominantly from deeper within the lungs and is expected to contain a higher concentration of virus particles. As such, based on the sites of severe infection in the respiratory tract, we hypothesize that the volume of air expelled by secondary and tertiary pulses could contain a higher viral load~(illustrated in Fig.~\ref{fig:motivation}) and that the resulting vortex-vortex interactions could significantly influence the dispersion of these more infectious particles.


It is now well established that interactions between turbulence and particles can give rise to preferential concentration, which describes the accumulation of particles away from highly vortical regions of the turbulent flow~\cite{eaton1994preferential, elghobashi1992direct, ireland2016effect1, salazar2008experimental, chun2005clustering}. When the Stokes number, defined as the ratio of particle-to-fluid time scales, is near unity, particles are directed by coherent vortical structures to create non-homogeneities in concentration and the onset of clusters. Large-scale velocity gradients present in free-shear flows affect the transport of small (Kolmogorov-scale) heavy particles and the clustering process at small scales~\citep{nicolai2013spatial,buchta2019sound}.  \citet{gualtieri2009anisotropic} showed that free shear flows generate anisotropic velocity fluctuations which, in turn, arrange particles in directionally biased clusters.  In the presence of gravity, preferential concentration by turbulence has been observed to cause particles to further accumulate near the downward moving side of vortices, referred to as preferential sweeping~\cite{maxey1987gravitational, wang1993settling, aliseda2002effect}. The gravitational settling of aerosol particles can be enhanced by this mechanism by as much as $50 \%$~\cite{wang1993settling}.

Turbulent transport in statistically stationary, axisymmetric, free jets has been well characterized experimentally~\cite{panchapakesan1993turbulence,hussein1994velocity,amielh1996velocity} and numerically~\cite{boersma1998numerical,wang2010direct,taub2013direct}. \citet{chein1987effects} demonstrated that particles with relatively small Stokes numbers disperse laterally at approximately the same rate as fluid particles, while particles with larger Stokes numbers exhibit significantly less dispersion.  In particular, particles with intermediate Stokes numbers are transported laterally farther than fluid particles due to enhanced entrainment by vortex structures. Shortly after, \citet{longmire1992structure} showed that particles become clustered in the saddle regions downstream of vortex rings and are propelled away from the jet axis by the outwardly moving flow. More recently, direct numerical simulations (DNS) of particle-laden round jets by \citet{li2011resolution} showed that all particles, regardless of their size, tend to preferentially accumulate in regions with larger-than-mean fluid streamwise velocity. Particle dispersion was found to be directly associated with three-dimensional vortex structures. While incredibly valuable, the aforementioned studies are restricted to jets with inflow characteristics that remain constant in time. By contrast, the transient characteristics of turbulent pulsatile jets are far less understood.

In this work, a realistic human cough is investigated computationally through DNS of pulsatile, turbulent, particle-laden jets. Fully-developed turbulence is provided at the orifice exit (mouth) using data obtained from an auxiliary simulation of turbulent pipe flow. The flow rate of the incoming turbulence is modulated in time according to a prescribed profile that controls the number of pulses, its duration, and peak amplitude. Particles are seeded in the flow with diameters sampled from a lognormal distribution informed by experimental measurements from the literature. Two-phase statistics, in particular fluid entrainment and particle evolution, are then reported for each case, with emphasis on the effect of pulsatility on the resulting vortex structures and particle dispersion.

\section{Simulation details}\label{sec:num}

\subsection{Flow configuration}
\begin{figure}[!h]
\captionsetup[subfigure]{labelformat=simple}
\centering
{\includegraphics[width=0.95\textwidth]{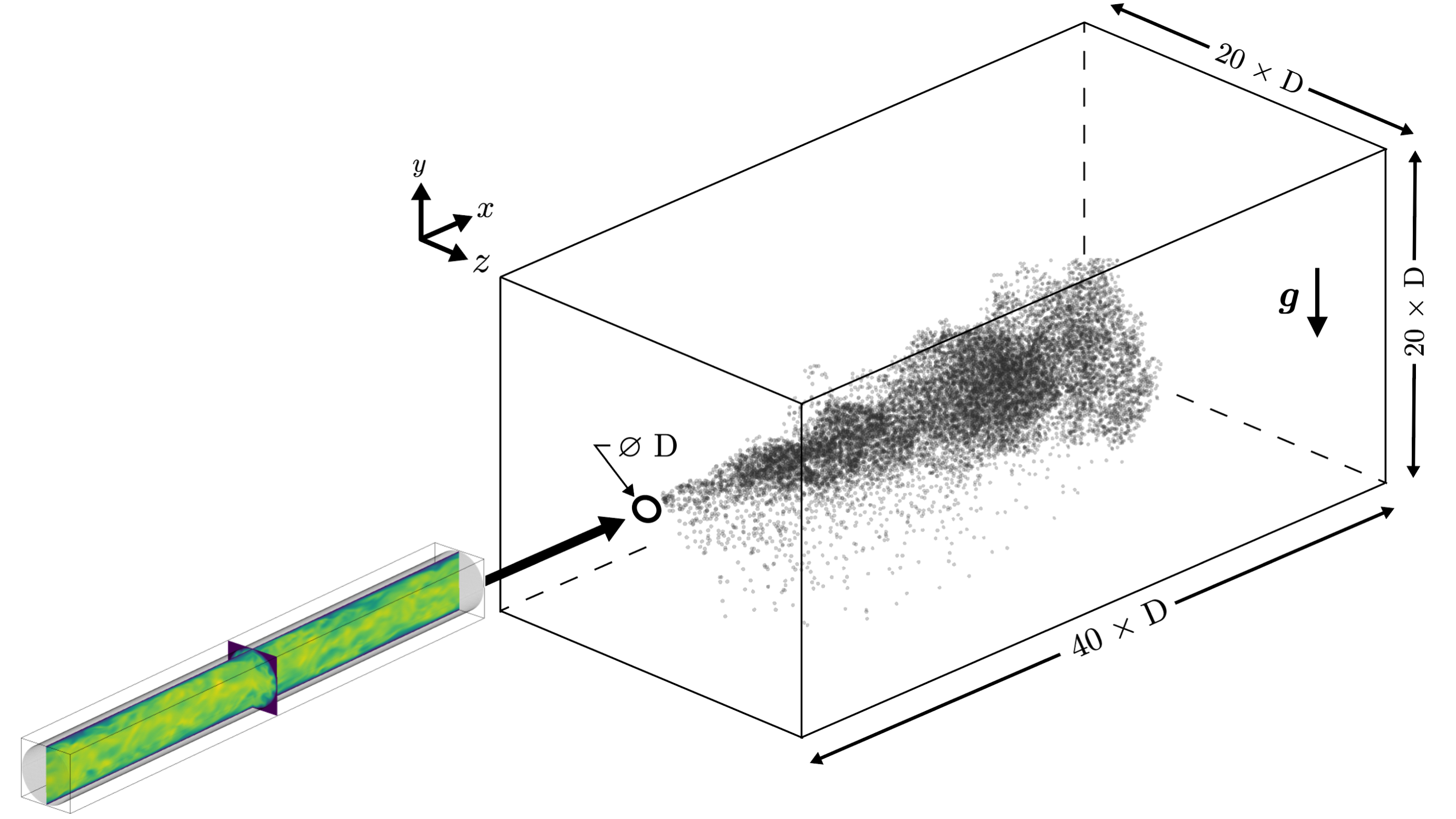}}
\caption{Sketch of the simulation setup showing the auxiliary pipe flow simulation providing fully-developed turbulence at the jet orifice. Particles are visualized at $t = 2$ s.}
\label{fig:domain}
\end{figure}

The present work considers a three-dimensional pulsatile jet laden with liquid droplets expelled into an ambient surrounding. Particles are considered to be well characterized as water droplets, and thus their density is held constant $\rho_p = 998$ kg/m$^3$. The fluid is considered to be air with a density of $\rho=1.172\,$kg/m$^3$ and kinematic viscosity of $\nu = 1.62\times 10^{-5}$m$^2$/s. The diameter of the orifice exit (mouth) is taken to be $D=0.02$ m. A Cartesian domain with length in the $x$ (streamwise), $y$ (spanwise, gravity-aligned) and $z$ (spanwise) directions are $L_x=40D$, $L_y=20D$, and $L_z=20D$, respectively (see Fig.~\ref{fig:domain}). The domain is discretized using $N_x=1024$ and $N_y=N_z=420$ grid points, with exponential grid stretching in the $y$- and $z$-directions. The spanwise grid spacing varies between $4.98\times10^{-4}$ m $\le \Delta y, \Delta z \le 2.1\times10^{-3}$ m such that the minimum grid spacing at the jet centerline is $D/40$. Previous work has shown this level of resolution is sufficient for free-shear jets at similar Reynolds numbers~\citep{li2011resolution}. A Dirichlet boundary condition is enforced at the jet inlet, a convective outflow is enforced at the downstream boundary, and all other boundaries are treated as slip walls.  To prevent fluid recirculation within the computational domain, a co-flow is introduced along the positive $x$-direction with velocity magnitude $0.32$ m/s. The co-flow is $\sim 7\%$ of the peak inflow velocity $U_0$ and was observed to have negligible effect on the particle dynamics. 

\subsection{Pulsatile inflow}\label {subsection:pulse}
Fully-developed turbulence is fed into the jet orifice using an auxiliary simulation of a turbulent pipe flow. The auxiliary simulation was performed using $256$ grid points across the diameter with a bulk velocity of $U_0=4$ m/s (a typical peak velocity associated with expiratory events~\cite{bourouiba2014violent,abkarian2020speech}), corresponding to a bulk Reynolds number ${\rm Re}_b=U_0D/\nu=4938$.  Further details on the pipe flow simulation are provided in Appendix~\ref{sec:pipe}.  Here, we note that the turbulent pipe simulation is statistically stationary and evolves to a constant bulk velocity $U_0$ defined by an imposed pressure gradient. To obtain a pulsatile turbulent inflow in the main simulation, the fluid velocity at the jet inlet $\bm{u}(x=0,y,z;t)$ is adjusted dynamically to control the volumetric flow rate in time. Building upon the experimental observations of \citet{gupta2009flow}, we propose a self-similar profile for the bulk velocity resulting from multiple expulsions.

\begin{figure}[!h]
\captionsetup[subfigure]{labelformat=simple}
\centering
{\includegraphics[width=0.55\textwidth]{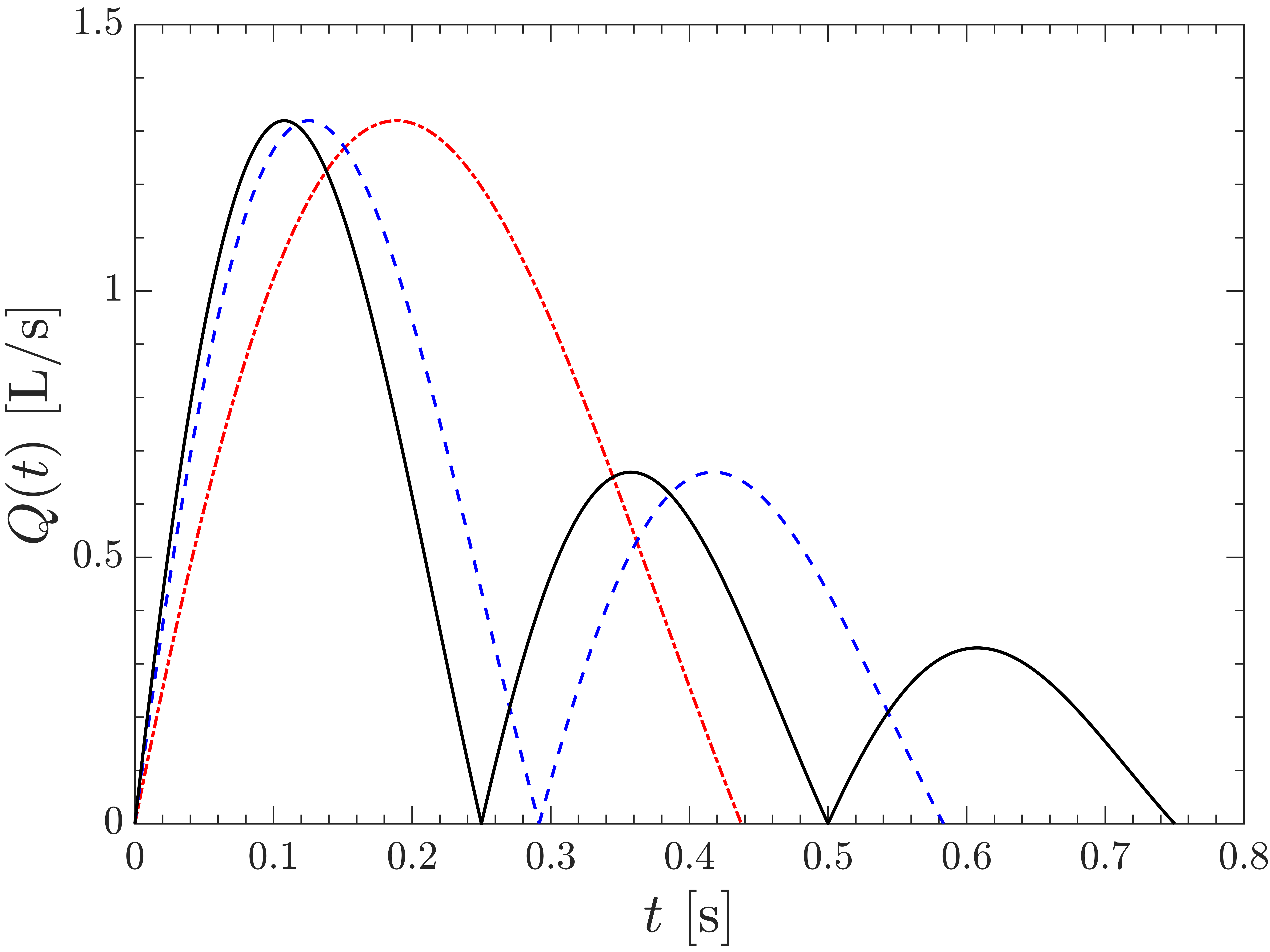}}
\caption{The simulated cough profiles used for the one pulse (\textcolor{red}{\bm{$- \cdot$}}), two pulse (\textcolor{blue}{\bm{$--$}}), and three pulse (\textcolor{black}{\bm{$-$}}) cases defined by Eq.~\eqref{damped_pulse}.}
\label{fig:pulse_profile}
\end{figure}

The proposed functional form for the pulsatile volumetric flow rate $Q(t)$ is given by a damped sine wave according to
\begin{equation}
\label{damped_pulse}
Q(t) = Q_0 \lvert e^{-t/ \tau} \sin(\omega t) \rvert,
\end{equation}
where $Q_0=U_0A_0$ and $A_0=\pi D^2/4$ is the area of the orifice exit. The fluid velocity is read in from the auxiliary pipe flow simulation and rescaled such that $\int\bm{u}(x=0,y,z,t)\cdot\bm{n}\,{\rm d}y{\rm d}z = Q(t)$, where $\bm{n}=[1,~0,~0]^{\sf T}$ is the outward surface normal. 
In the present study, we consider three profiles corresponding to one, two, and three pulses (see Fig.~\ref{fig:pulse_profile}). The relaxation time is chosen to be $\tau = [0.63, 0.42, 0.36]\,$s, and the frequency is $\omega = [7.18, 10.77, 12.57] \,{\rm s}^{-1}$. For each case, the maximum velocity of exhaled airflow occurs at approximately 100 ms, consistent with measurements of coughing from human subjects~\citep{gupta2009flow}. The total duration of each profile varies but the inputs are defined to yield the same volume of expelled air so a fair comparison can be drawn between cases with different numbers of pulses. 

\subsection{Particle injection}
\begin{figure}[!h]
\captionsetup[subfigure]{labelformat=simple}
\centering
{\includegraphics[width=0.65\textwidth]{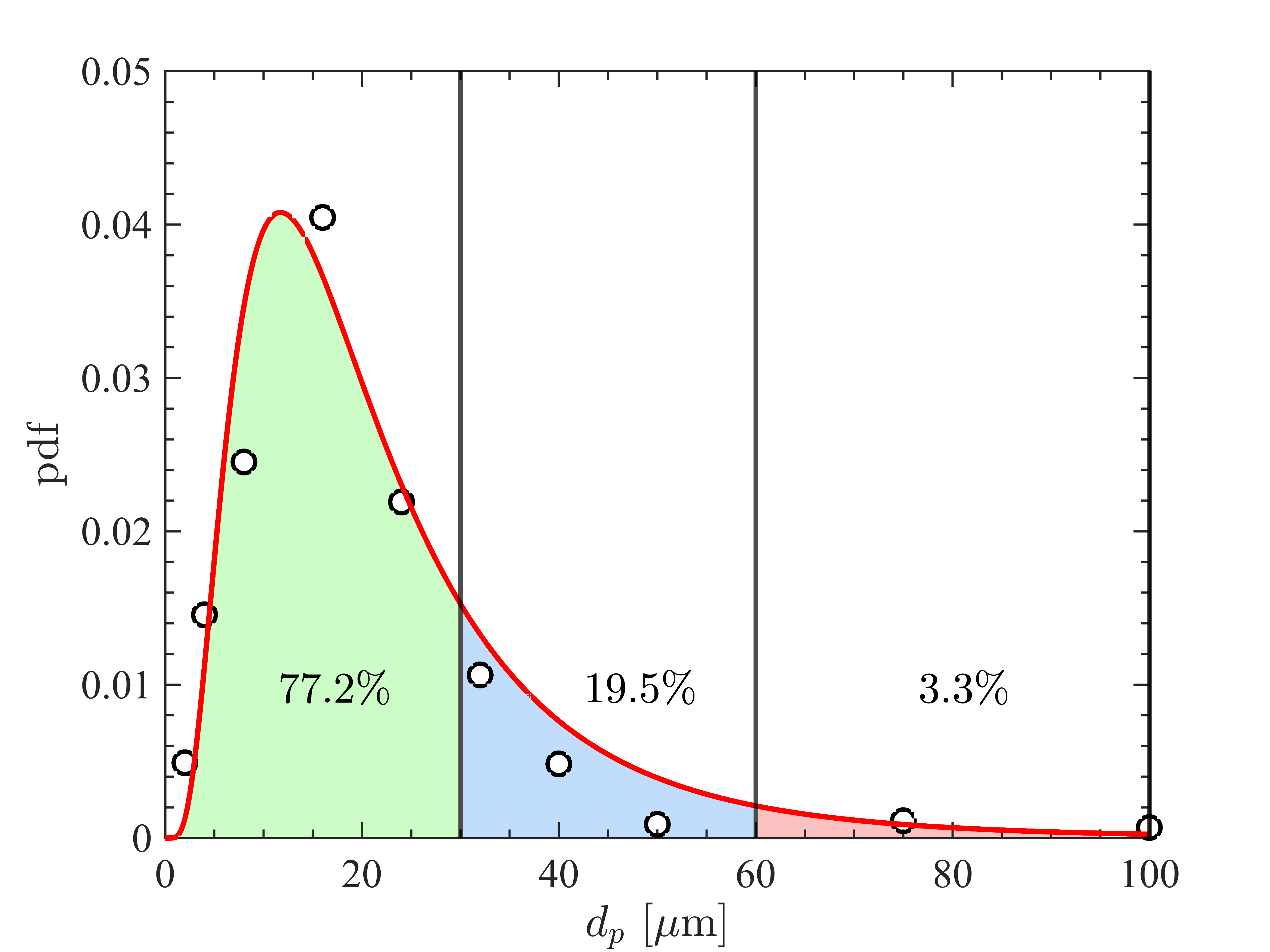}}
\caption{Lognormal size distribution used to sample particle diameters in the DNS for the pulsatile jet (\textcolor{red}{\bm{$-$}}) and experimental measurements ($\Circle$) by \citet{duguid1946} for droplets generated in realistic coughs. Color shading denotes three size ranges corresponding to small (green) intermediate (blue) and high (red) Stokes numbers.}
\label{fig:part_distro}
\end{figure}

To accurately characterize the particle size distribution generated by coughing, we employ a lognormal distribution fit to the experimental measurements of~\citet{duguid1946}. The particle diameter ranges between $1\le d_p\le100$ $\upmu$m with a mean of $24$ $\upmu$m and a standard deviation of $17.9$ $\upmu$m as shown in Fig.~\ref{fig:part_distro}. At each simulation timestep, particles are introduced at the inflow plane by assigning them a random position within the orifice and a diameter that is sampled from the aforementioned lognormal distribution. The number of particles per timestep is adjusted dynamically to achieve the same mass flow rate used for the fluid. Given the prescribed expulsion volume and particle size distribution, approximately 15,000 particles are generated at the end of a coughing spell,  representative of the typical quantity observed in experiments~\cite{lindsley2012quantitysize}.

The turbulence Stokes number, St$_\eta=\tau_{p} / \tau_{\eta}$, may be utilized to gauge the role of particle inertia, where $\tau_{p}=\rho_pd_p^2/(18 \rho \nu)$ is the particle response time, $\tau_{\eta} = (\nu/\varepsilon)^{1/2}$ is the Kolmogorov time scale of the fluid phase, and $\varepsilon$ is the turbulence dissipation rate. When analyzing the results in Sec.~\ref{sec:results}, particles are demarcated into three size ranges:  $d_p \in [1, 30]$, $[30, 60]$, and $[60, 100]~\upmu$m, which yields the following Stokes number ranges St $\in [0.0054, 4.87]$, $[4.87, 19.50]$, and $[19.50, 54.16]$. We note that the Stokes numbers are defined using values taken at the orifice exit and therefore will decay as particles evolve in time and downstream. Nevertheless, these Stokes number ranges provide insight into the relative inertia of the fluid and particle phases. Specifically, particles will behave ballistically in the large Stokes limit but act as fluid tracers in the small Stokes limit.

In real expiratory events, exhaled particles will exhibit a distribution of velocities that may deviate from the local air flow due to complex interactions in the upper respiratory tract. Motivated by the fact that the majority of particles lie within the first size bin, where Stokes numbers are small St $\in [0.0054, 4.87]$, we treat the particles as fluid tracers at the orifice exit and specify their initial velocity to be the fluid velocity interpolated at the particle position. Further details are provided in Appendix~\ref{sec:pipe}. We emphasize that this assumption of zero interphase slip velocity at the orifice exit is a significant assumption made within the present work. Future experimental studies will be required to find the extent at which this assumption can be considered valid.

\subsection{Governing equations}
The simulations are solved in Eulerian--Lagrangian framework, where individual particles are treated in a Lagrangian manner, and the gas phase is solved on a background Eulerian mesh. Due to the low concentrations considered in this study, volume fraction effects and two-way coupling between the phases are neglected. The governing equations for the incompressible carrier phase are given by
\begin{equation}
\label{continuity}
\nabla\cdot\bm{u}=0,
\end{equation}
and
\begin{equation}
\label{velocity}
\frac{\partial\bm{u}}{\partial t}+ \bm{u}\cdot\nabla\bm{u}=-\frac{1}{\rho}\nabla p +\nu\nabla^2\bm{u}+\bm{g},
\end{equation}
where  $\bm{u}=[u,~v,~w]^{\sf T}$ is the fluid velocity, $p$ is the hydrodynamic pressure, and $\bm{g}=[0, -g, 0]^{\sf T}$ is the gravitational acceleration with $g=9.8\,$m/s$^2$. The equations are implemented in the framework of the NGA code~\citep{desjardins2008high}. The Navier--Stokes equations are solved on a staggered grid with second-order spatial accuracy for both convective and viscous terms, and the semi-implicit Crank-Nicolson scheme is used for time advancement maintaining overall second-order accuracy.

Particles are treated in a Lagrangian manner where the translational motion of an individual particle `$i$' is given by
\begin{equation}
\frac{{\rm d}\bm{x}_p^{(i)}}{{\rm d}t} = \bm{v}_p^{(i)},
\end{equation}
\begin{equation} \label{newtonlaw}
m_p\frac{{\rm d}\bm{v}_p^{(i)}}{{\rm d}t}=\rho V_p \nabla \cdot \boldsymbol{\tau}[ \bm{x}_p^{(i)} ]+\bm{f}_{\text{drag}}^{(i)}+m_p \bm{g} ,
\end{equation}
where $\bm{x}_p^{(i)}$ and $\bm{v}_p^{(i)} = [u_p^{(i)},~v_p^{(i)},~w_p^{(i)}]^{\sf T}$ are the instantaneous particle position and velocity, respectively, and $m_p=\rho_p\pi d_p^3/6$ is the particle mass. Here $\bm{\tau}[\bm{x}_p^{(i)}]$ is the resolved fluid stress at the particle location and $\bm{f}_{\text{drag}}^{(i)}$ accounts for unresolved stress due to drag. In this work, the classic Schiller and Naumann drag correlation~\cite{clift2005bubbles} is used to account for finite Reynolds number effects, given by
\begin{equation}\label{part_source}
\frac{\bm{\bm{f}}_{\text{drag}}^{(i)}}{m_p}=\frac{1+0.15 \text{Re}_p^{0.687}}{\tau_p}\left(\bm{u}[\bm{x}_p^{(i)}]-\bm{v}_p^{(i)}\right),
\end{equation}
where $\bm{u}[\bm{x}_p^{(i)}]$ is the fluid velocity at the location of particle `$i$' and $\text{Re}_p =\| \bm{u}[\bm{x}_p^{(i)}]-\bm{v}_p^{(i)}\| d_p/\nu$ is the particle Reynolds number. The particle equations are advanced in time using a second-order Runge--Kutta scheme.

We briefly note that the present work does not consider thermodynamic effects, such as evaporation and buoyancy, in order to isolate the role of pulsatility on particle dispersion and minimize the parameter space under study. Recent experiments showed that thermal effects are small until the jet speeds are reduced to ambient speeds~\cite{abkarian2020speech}. Thus, the present work focuses on the near-mouth region, where the unsteadiness of the expiratory events is expected to have a more pressing role on particle dynamics. 


\section{Results and discussion}\label{sec:results}

\subsection{Pulsatile free-shear jet}\label{sec:jet-results}

\subsubsection{Flow visualization}
\begin{figure}[h!]
\begin{center}
\begin{subfigure}[t]{0.7\textwidth}
\includegraphics[width=\textwidth]{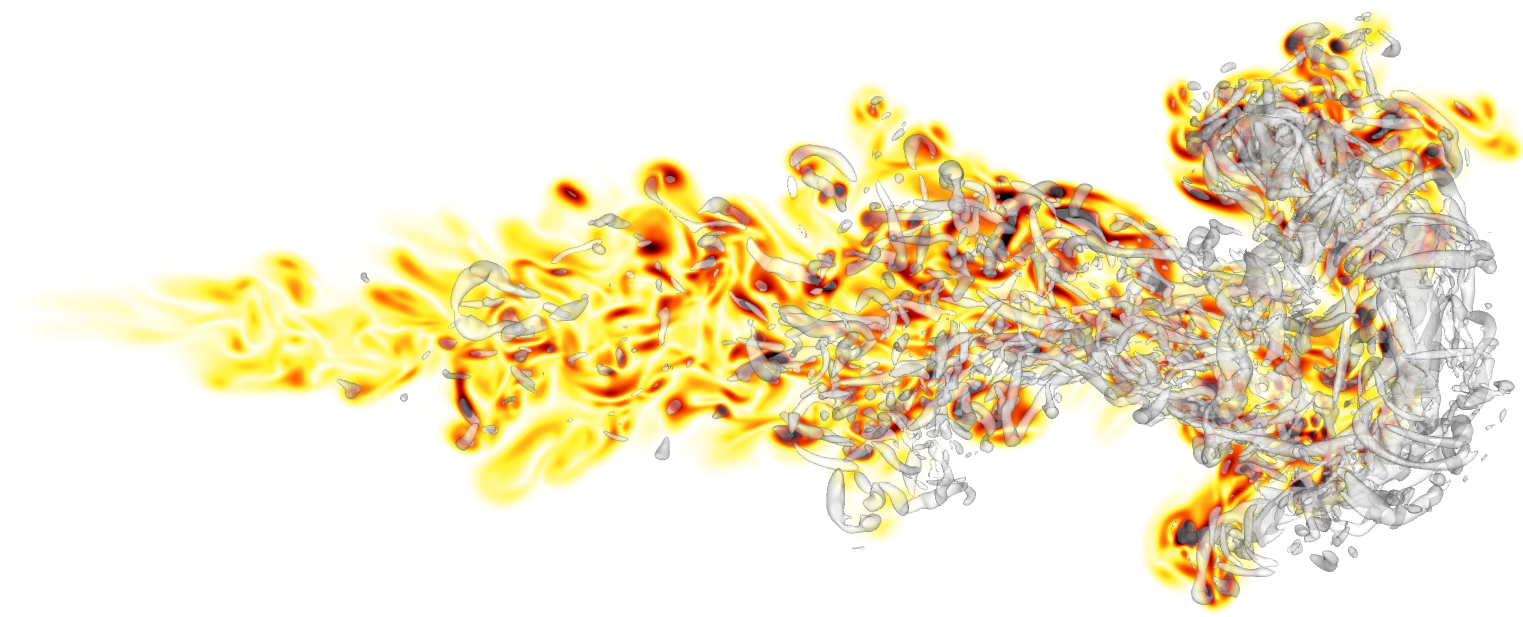}
\caption{Single-pulse}
\label{fig:Q1}
\end{subfigure}
\begin{subfigure}[t]{0.7\textwidth}
\includegraphics[width=\textwidth]{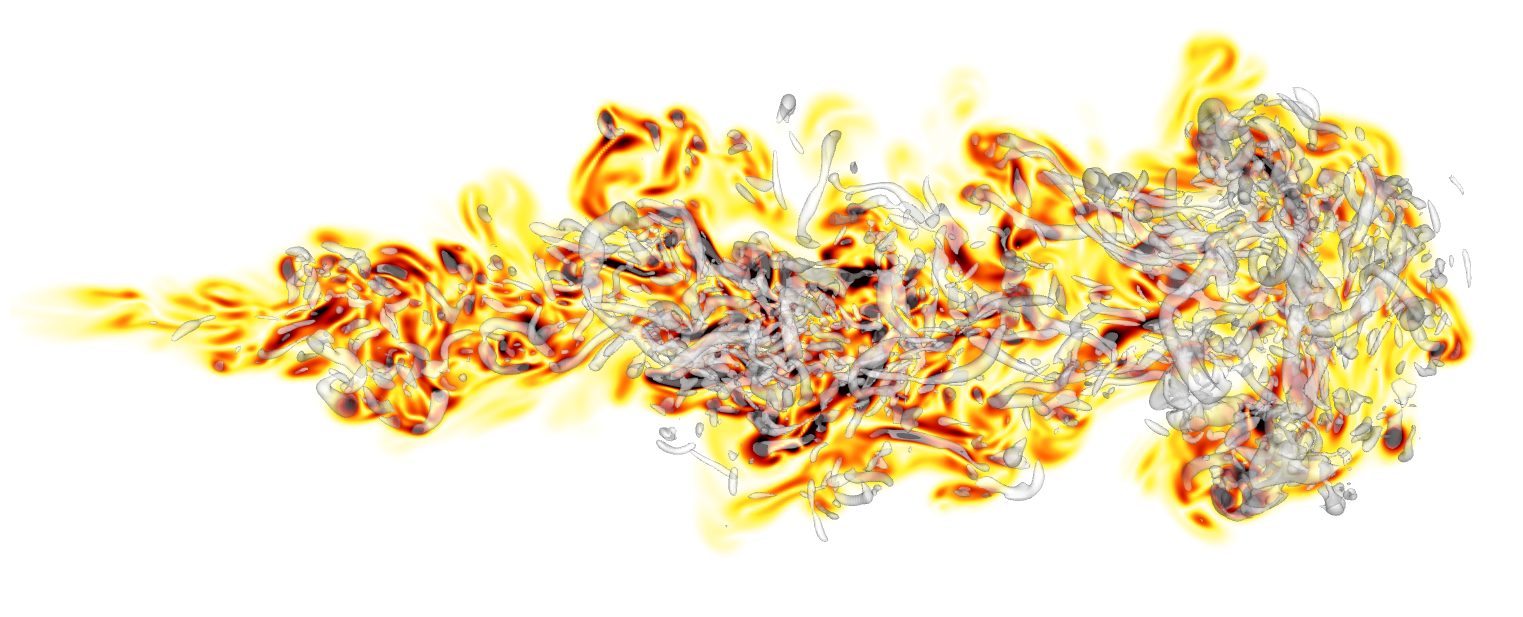}
\caption{Two-pulse}
\label{fig:Q2}
\end{subfigure}
\begin{subfigure}[t]{0.7\textwidth}
\includegraphics[width=\textwidth]{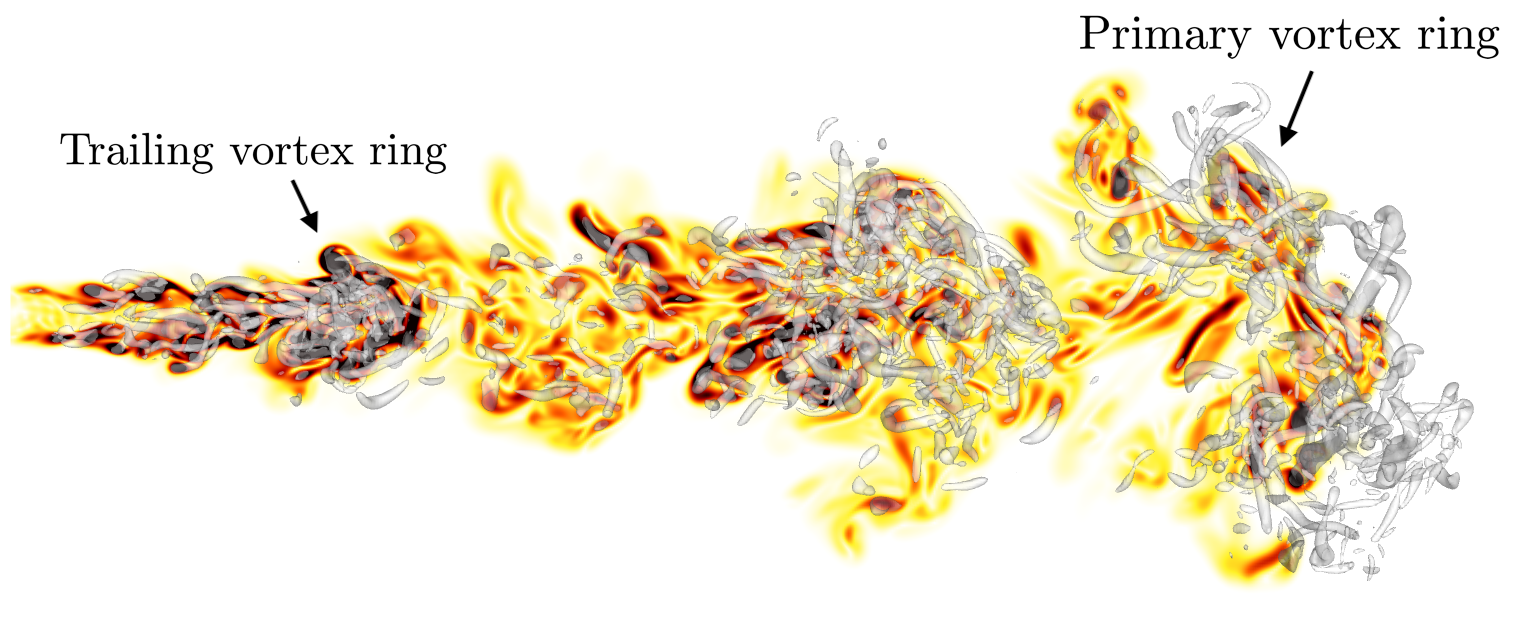}
\caption{Three-pulse}
\label{fig:Q3}
\end{subfigure}
\caption{Instantaneous snapshots of the turbulent pulsatile jet at $t=0.75$ s. Color shows vorticity magnitude varying from 0 (white) to $0.75 \lVert\bm{\omega}\rVert D / U_0$ (black). Iso-surface of positive $Q$-criterion shown in gray.}
\label{fig:Qcrit}
\end{center}
\end{figure}

Figure~\ref{fig:Qcrit} shows visualizations of the single-, two-, and three-pulse jets at $t=0.75$ s, immediately after the final pulse is complete (cf., Fig.~\ref{fig:pulse_profile}). Inspection of the vorticity magnitude $\norm{\bm{\omega}}$, with $\bm{\omega}=[\omega_x,~\omega_y,~\omega_z]^{\sf T}$, reveals distinct differences between the three cases. Vortical structures are visualized using the $Q$-criterion~\citep{hunt1988eddies}, defined as the second invariant of the velocity gradient tensor, given by
\begin{equation}\label{eq:Qcrit}
Q_{\rm crit}=\frac{1}{2}\left(\norm{\bm{\Omega}}^2-\norm{\bm{S}}^2\right)>0,
\end{equation}
where $\bm{\Omega}=(\nabla\bm{u}-\nabla\bm{u}^{\sf T})/2$ and $\bm{S}=(\nabla\bm{u}+\nabla\bm{u}^{\sf T})/2$ are the antisymmetric and symmetric components of the velocity gradient, respectively. Physically speaking, the $Q$-criterion represents a local balance between shear strain and vorticity, with vortices being defined by regions where rigid body rotation is greater than the rate-of-strain.

To demonstrate the effect of pulsatility on the fluid phase, we first consider the location of vortical structures at the end of the third pulse (see Fig. \ref{fig:Qcrit}). For the single-pulse case, a primary vortex ring structure is generated at the downstream edge of the jet while vorticity is minimal near the orifice exit. By contrast, the two-pulse and three-pulse cases exhibit multiple vortex ring structures, corresponding to the number of pulses, with comparatively higher regions of vorticity upstream near the orifice. Vortical structures in the near-orifice region, which are absent from the single-pulse case, will impact the transport of low inertia particles. Specifically, the higher vorticity levels observed in two- and three-pulse cases are expected to accelerate and entrain late-stage injected particles to a larger degree when compared to the single-pulse case. However, the strength of the leading vortex structure for two- and three-pulse cases is significantly attenuated from the single-pulse case. The role of pulsatility on particle dynamics is reserved for \S~\ref{sec:part-results}.

\subsubsection{Entrainment}
Entrainment of the surrounding air into the jet plays a key role on its transport properties. In the seminal paper by \citet{morton1956turbulent}, it was suggested that entrainment, defined as the mean radial velocity at the edge of an axisymmetric boundary layer (in this case edge of the jet), is proportional to the axial velocity, i.e., $\langle u_r\rangle=\alpha\langle u\rangle$, 
where $u_r=(vy+wz)/r$ is the radial velocity with $r=\sqrt{y^2+z^2}$ the radial position. Due to the lack of statistical stationarity in the present configuration, angled brackets used herein denote an average in the circumferential direction but not in time. \citet{pham2006large} suggested that the coefficient of entrainment, $\alpha$, can be estimated as
\begin{equation}
\alpha(x, t)=\frac{\frac{{\rm d}}{{\rm d} x} \int_{0}^{\infty}\langle u\rangle r\,{\rm d} r}{\delta(x,t) \langle u\rangle \vert_{r=0}},
\end{equation}
where $\delta$ is the momentum balance length scale defined as
\begin{equation}
\delta^{2}(x,t)=\frac{2\left(\int_{0}^{\infty}\left\langle u \right\rangle r\,{\rm d} r\right)^{2}}{\int_{0}^{\infty}\left\langle u \right\rangle^{2} r\,{\rm d} r}.
\end{equation}
\citet{taub2013direct} showed that $\alpha$ is approximately constant when the jet has achieved self similarity. For the pulsatile transient jets considered here,  the centerline velocity $\langle u \rangle \vert_{r=0}$ is zero near the orifice when the pulses complete, resulting in an ill-defined $\alpha$. To this end, we propose an alternative definition based on the jet bulk velocity $U_0$ according to
\begin{equation}
\alpha(x, t)=\frac{\frac{{\rm d}}{{\rm d} x} \int_{0}^{\infty}\langle u\rangle r\,{\rm d} r}{\delta(x,t)\,U_0}.
\end{equation}

\begin{figure}[h!]
\begin{center}
\begin{subfigure}[t]{0.329\textwidth}
\includegraphics[width=\textwidth]{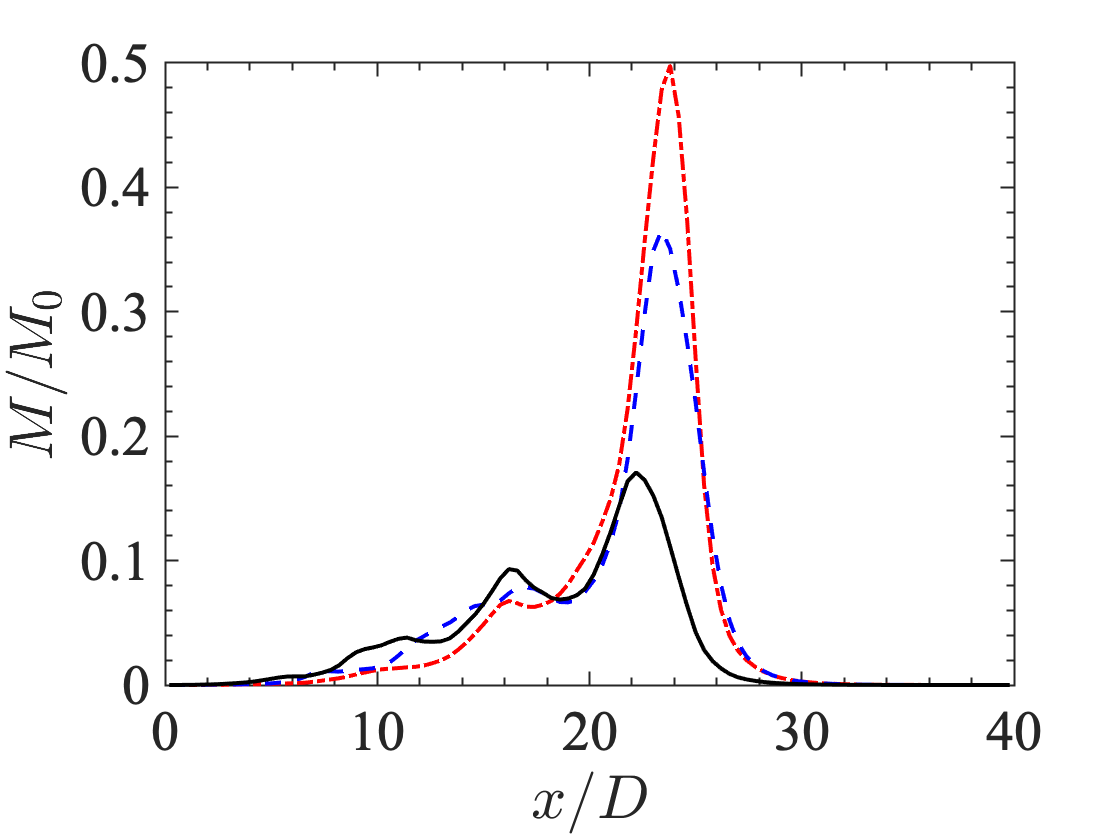}
\caption{$t = 1.0$ s}
\label{fig:m_t1}
\end{subfigure}
\begin{subfigure}[t]{0.329\textwidth}
\includegraphics[width=\textwidth]{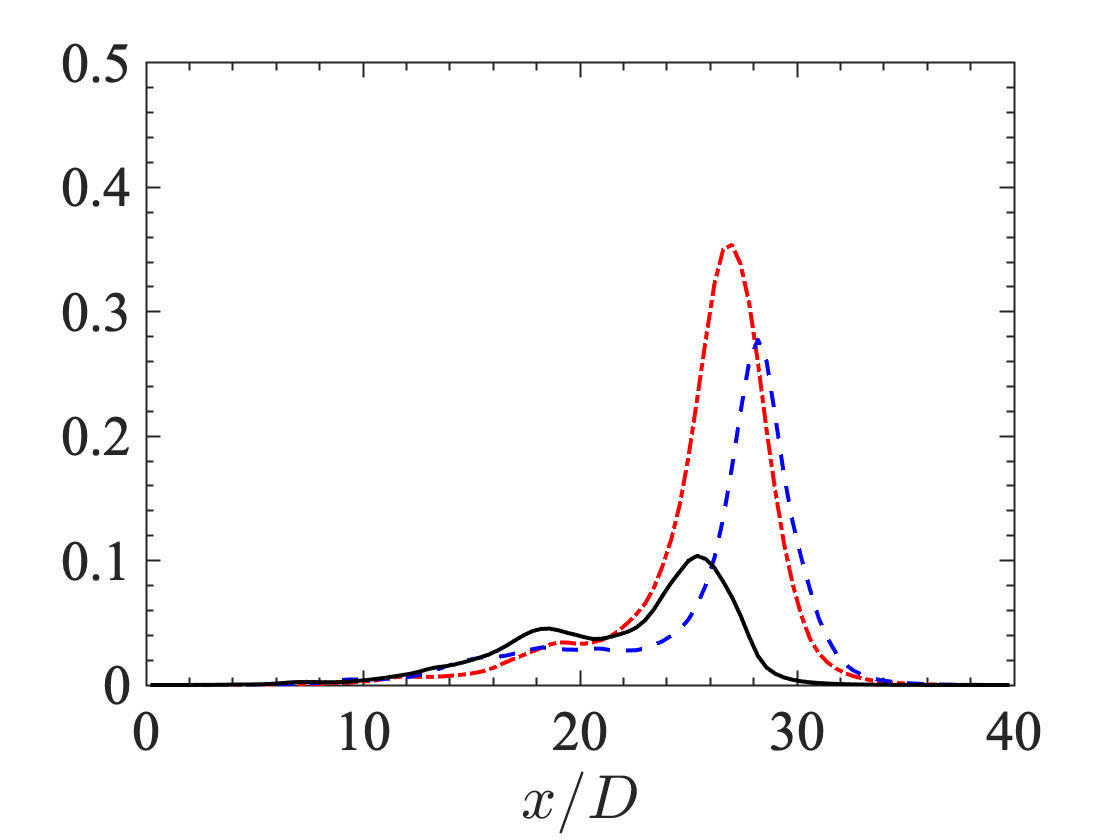}
\caption{$t = 1.5$ s}
\label{fig:m_t2}
\end{subfigure}
\begin{subfigure}[t]{0.329\textwidth}
\includegraphics[width=\textwidth]{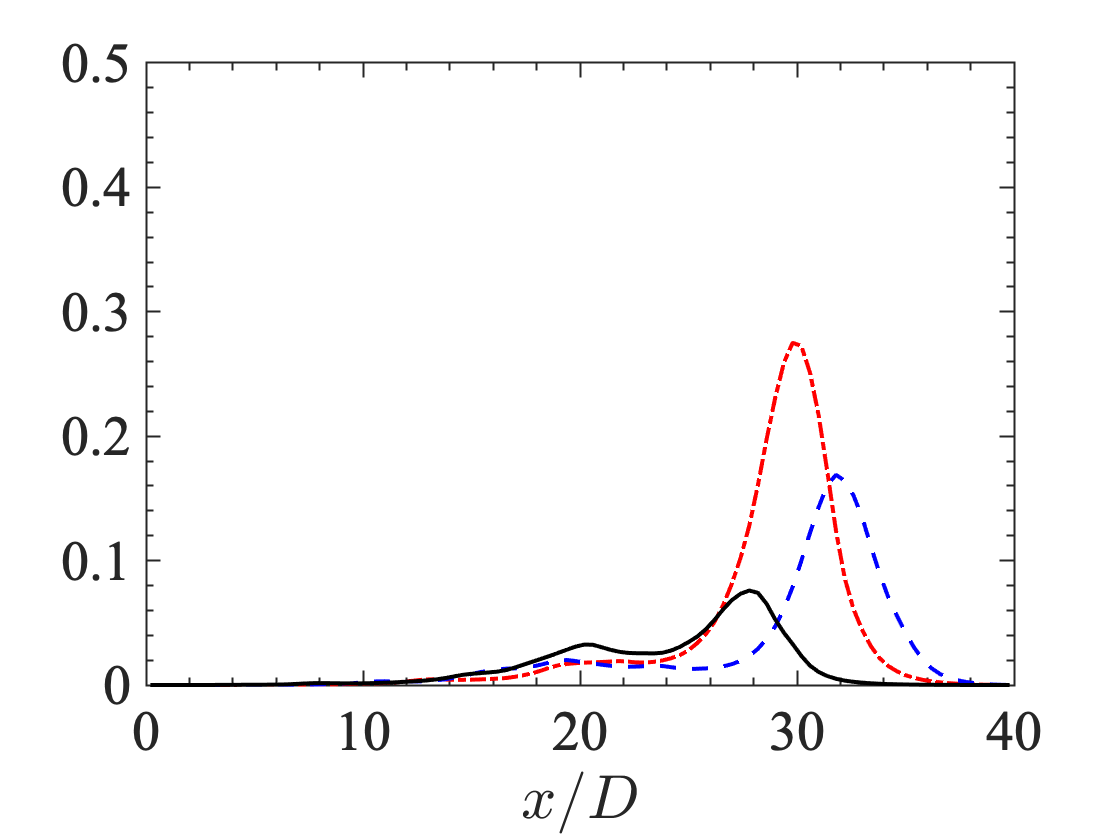}
\caption{$t = 2.0$ s}
\label{fig:m_t3}
\end{subfigure}
\begin{subfigure}[t]{0.329\textwidth}
\includegraphics[width=\textwidth]{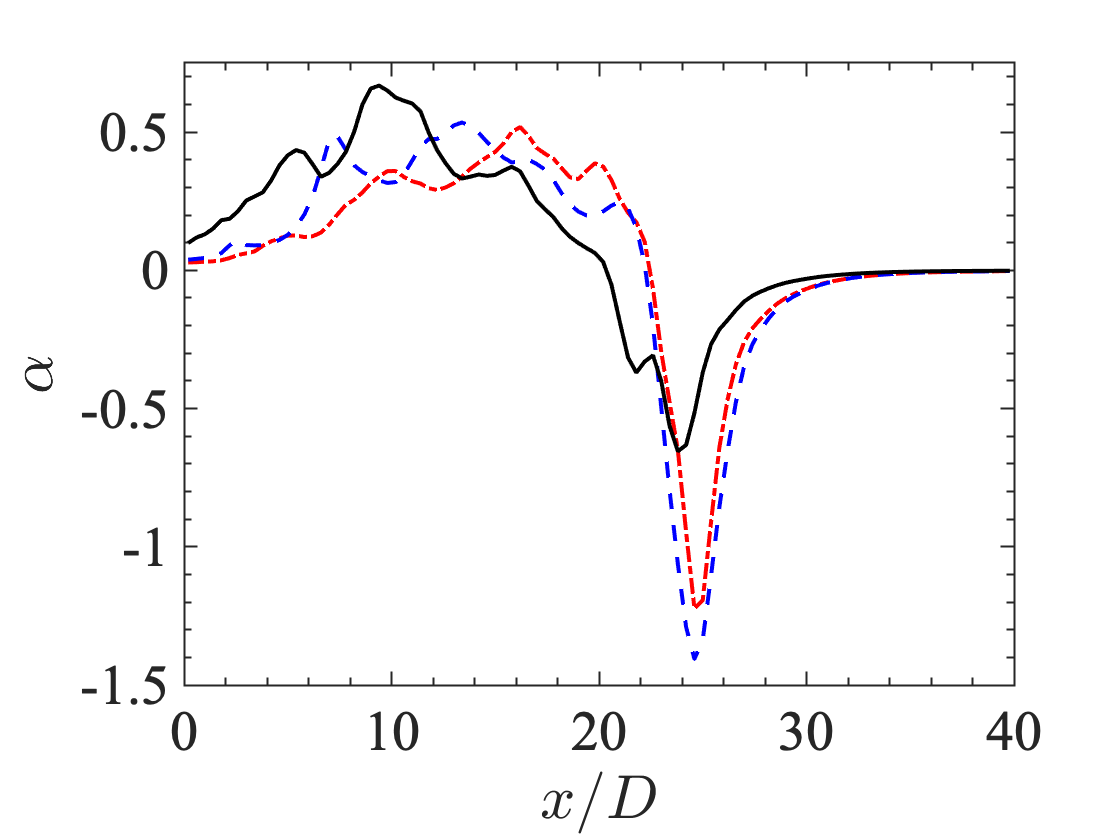}
\caption{$t = 1.0$ s}
\label{fig:alpha_t1}
\end{subfigure}
\begin{subfigure}[t]{0.329\textwidth}
\includegraphics[width=\textwidth]{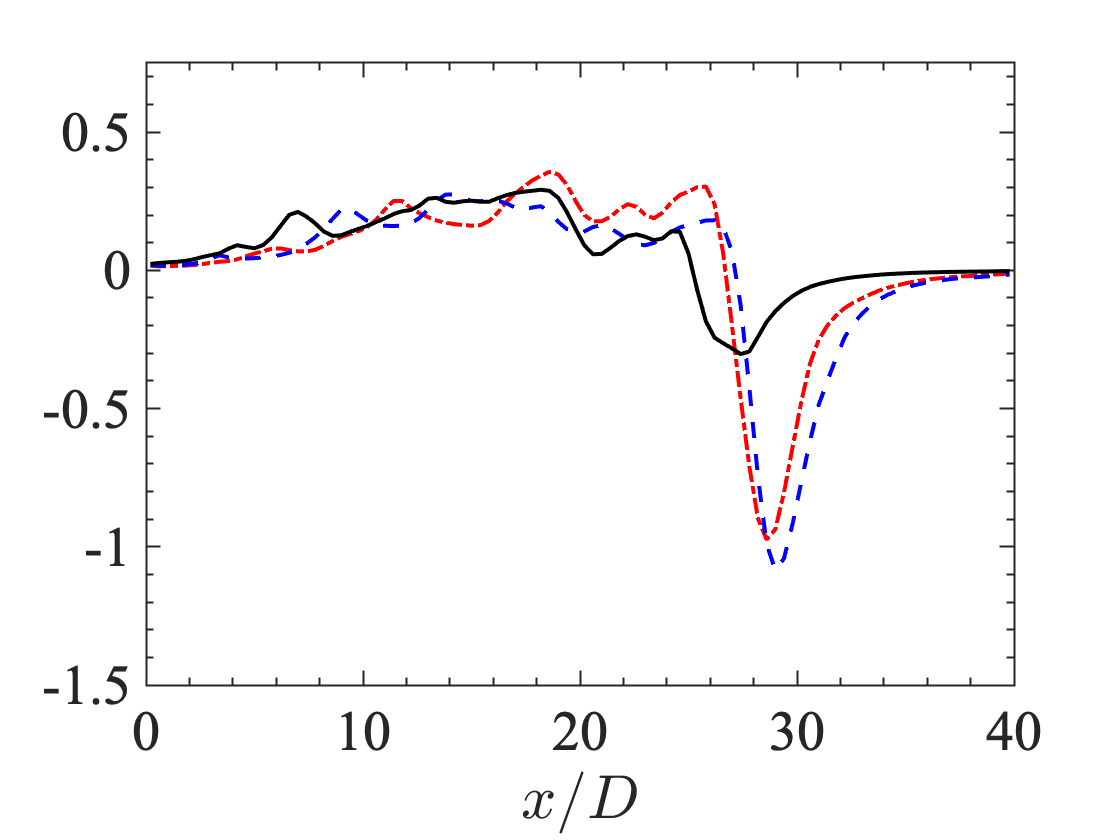}
\caption{$t = 1.5$ s}
\label{fig:alpha_t2}
\end{subfigure}
\begin{subfigure}[t]{0.329\textwidth}
\includegraphics[width=\textwidth]{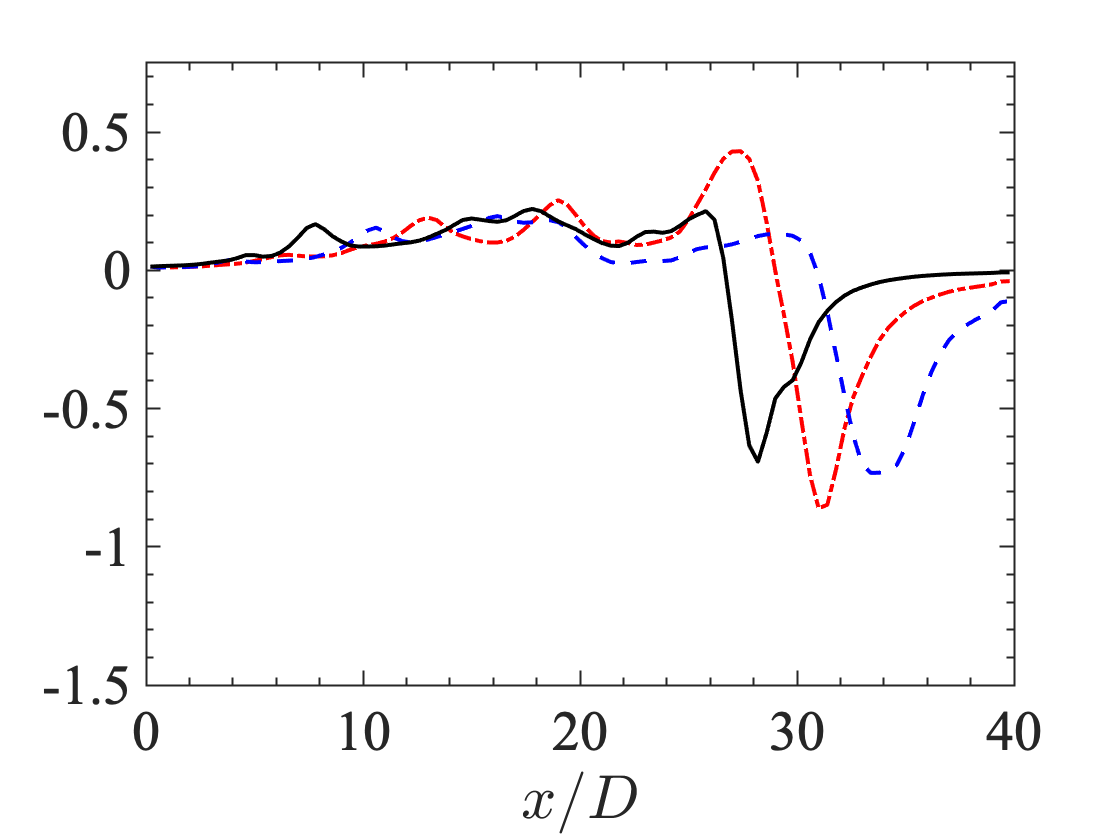}
\caption{$t = 2.0$ s}
\label{fig:alpha_t3}
\end{subfigure}
\caption{Top: specific momentum $M/M_0$; and bottom: entrainment coefficient $\alpha$ as a function of streamwise location $x/D$ at $t = 1.0\,{\rm s}, 1.5\,{\rm s,\ and\ } 2.0\,{\rm s}$ for the three different cases. Line types and colors same as Fig.~\ref{fig:pulse_profile}.}
\label{fig:entrain_coeff}
\end{center}
\end{figure}

The entrainment coefficient $\alpha$ as a function of streamwise location $x/D$ at $t = 1$ s, $1.5$ s, and $2$ s for each case is summarized in Fig.~\ref{fig:entrain_coeff}. The corresponding specific momentum $M=\int_{0}^{\infty}\left\langle u \right\rangle^{2} r\,{\rm d} r$, normalized by the maximum value $M_0 = \int_{0}^{\infty} U_0^{2} r\,{\rm d} r = 8 \times 10^{-4}\,{\rm m}^4/{\rm s}^2$, is also reported to indicate the instantaneous location of each pulse. It can be seen that the single-pulse case generates significantly more momentum downstream at the jet front, while the two-pulse and three-pulse cases exhibit a wider distribution in momentum in the streamwise direction. In addition, the momentum profile is bi-modal for the two-pulse case and tri-modal for the three-pulse case at $t = 1$ s, where each peak coincides with the peak amplitude of each pulse. As a result, larger values of momentum are observed near the orifice for the two-pulse and three-pulse cases, which proceed to decay as the flow propagates downstream.

The entrainment coefficient is seen to be positive for all three cases at $x/D<20$ when $t=1$ s, which indicates a net entrainment of ambient air into the jet. Beyond this point ($x/D>20$) $\alpha$ becomes negative with the largest magnitude in concert with the first pulse. Note that unlike in the more traditional statistically stationary jet, where $\alpha$ is positive for all $x$, here the primary vortex ring generated by the first pulse induces a net momentum flux from the jet into the ambient air, resulting in $\alpha<0$. By comparing the three cases at $t = 1$ s, it is observed that $\alpha$ is larger in the upstream regions ($x/D<16$) for the multi-pulse cases compared to the single-pulse case due to the vortical structures generated by subsequent pulses. In addition, the three-pulse case exhibits significantly larger entrainment at the jet front where the primary vortex ring resides.

Pulsatility is also observed to affect late-time single-phase dynamics of the jet. By comparing Figs.~\ref{fig:m_t1}--\ref{fig:m_t3}, the primary vortex of the two-pulse jet is seen to travel at a higher speed, followed by the single-pulse case then the three-pulse case. This results in noticeable differences in the entrainment coefficient between each case at late times. The following sections seek to understand how these  combined effects influence particle entrainment and dispersion.

\subsection{Role of pulsatility on particle dispersion}\label{sec:part-results}
\begin{figure}[!h]
\captionsetup[subfigure]{labelformat=simple}
\centering
{\includegraphics[width=\textwidth]{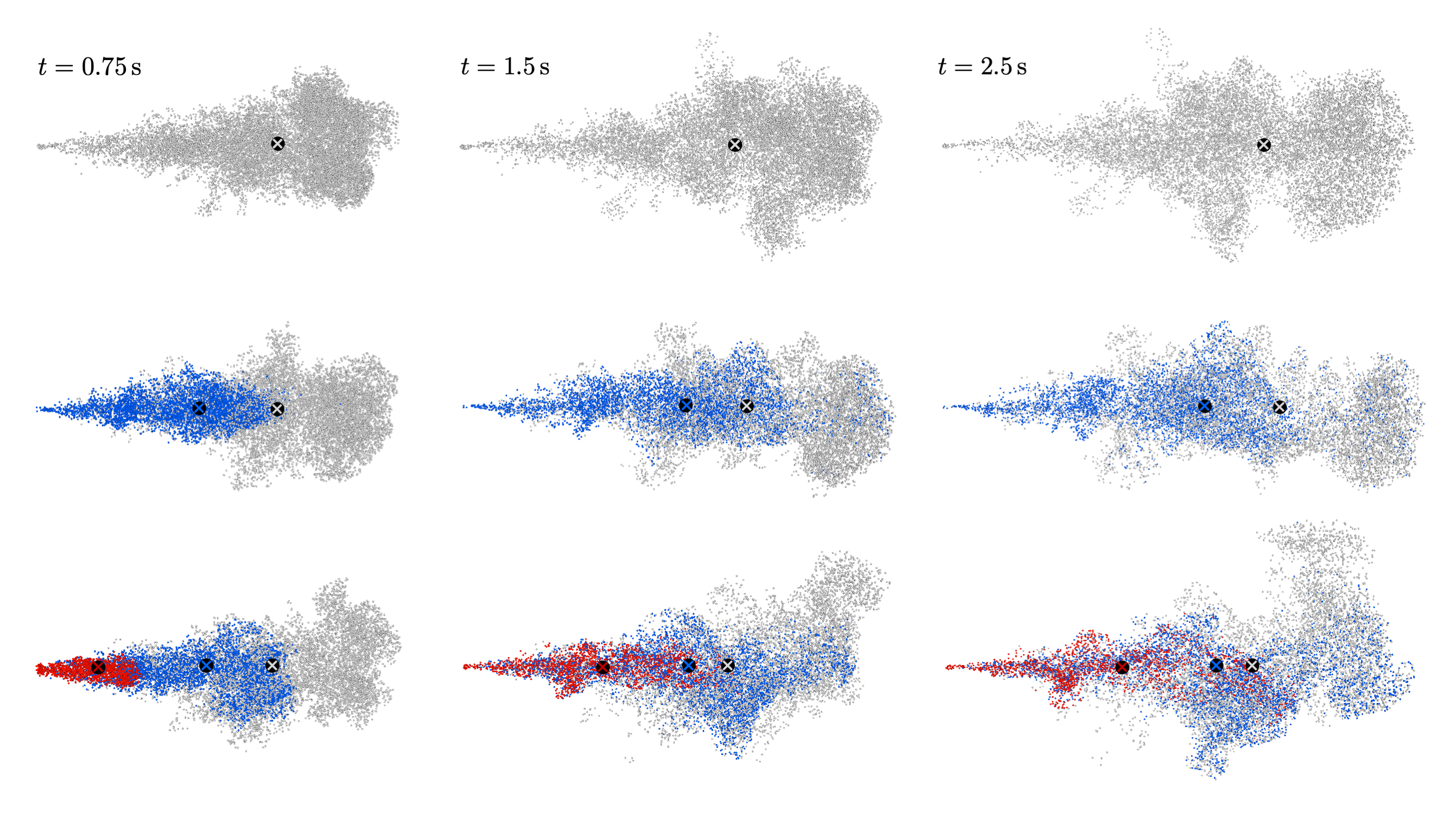}}
\caption{Particle distribution for the one-pulse (top), two-pulse (middle), and three-pulse (bottom) case at different times ($t=0.75, 1.5, {\rm \ and \ } 2.5 \,{\rm s}$) shown from the top ($x-z$ plane). Particles originating from the first, second, and third pulse are colored gray, blue, and red, respectively, with $\otimes$ denoting the geometric center of the cloud associated with each pulse.}
\label{fig:part_topview}
\end{figure}

Visualizations of the particle cloud at three instances in time are shown in Fig.~\ref{fig:part_topview}. Particles are colored by the corresponding pulse they were injected into. At each timestep, the total number of particles associated with each pule, $N_p(t)$, is identified and used to define the geometric center of the cloud according to $\bm{x}_c=\sum_{i=1}^{N_p}\bm{x}_p^{(i)}/N_p$. Qualitative differences in the cloud evolution can be seen between the three cases. Careful inspection of Fig.~\ref{fig:part_topview} reveals that the overall penetration of the cloud is slightly hindered with increased pulsatility, although this effect is minor. More pronounced is the penetration of particles associated with subsequent pulses. This is best seen in the three-pulse case, where particles emanating from the second pulse (colored blue) penetrate to the cloud front when $t>1.5$ s, despite being injected with lower velocity. In addition, particles from the third pulse (colored red) in the three-pulse case penetrate nearly as far as particles from the second pulse in the two-pulse case. In summary, the geometric centers associated with particles of later pulses travel further downstream with increased pulsatility, and advance upon particles injected earlier. The observed increase in penetration by secondary and tertiary pulses has important connotations as it is expected that later expulsions could contain higher viral concentrations depending on the location in the respiratory tract where the infection resides. Therefore, the aforementioned phenomena may prove to be significant when determining distances at which an infectious person can be harmful to others; as the enhanced transport of high viral load droplets is expected to increase the probability of transmission.



\begin{figure}[!h]
\captionsetup[subfigure]{labelformat=simple}
\centering
{\includegraphics[width=0.85\textwidth]{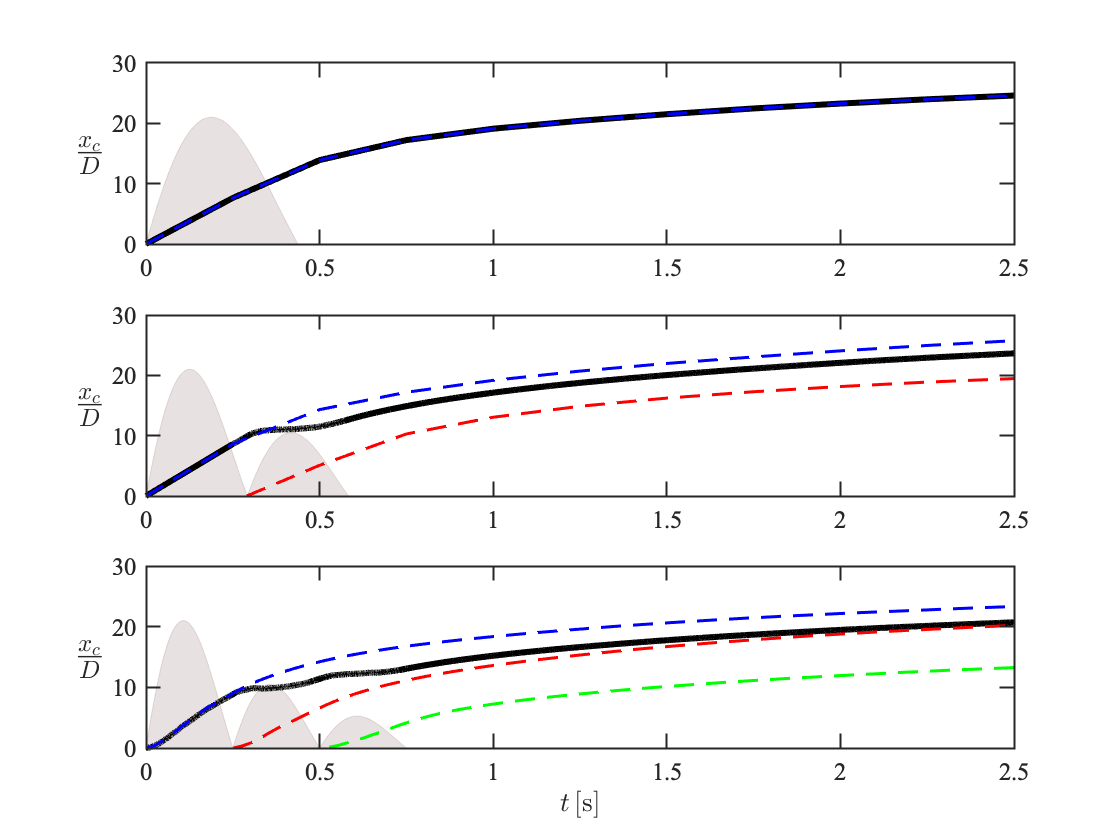}}
\caption{Temporal evolution of the geometric center of the overall cloud (\textcolor{black}{\bm{$-$}}) and first (\textcolor{blue}{\bm{$--$}}), second (\textcolor{red}{\bm{$--$}}), and third pulse (\textcolor{green}{\bm{$--$}}) in the $x$-direction for the one pulse (top), two pulse (middle), and three-pulse (bottom) cases. The flow rate $Q(t)$ associated with each pulse is shown in gray.}
\label{fig:x_penetration}
\end{figure}

\begin{figure}[!h]
\captionsetup[subfigure]{labelformat=simple}
\centering
{\includegraphics[width=0.85\textwidth]{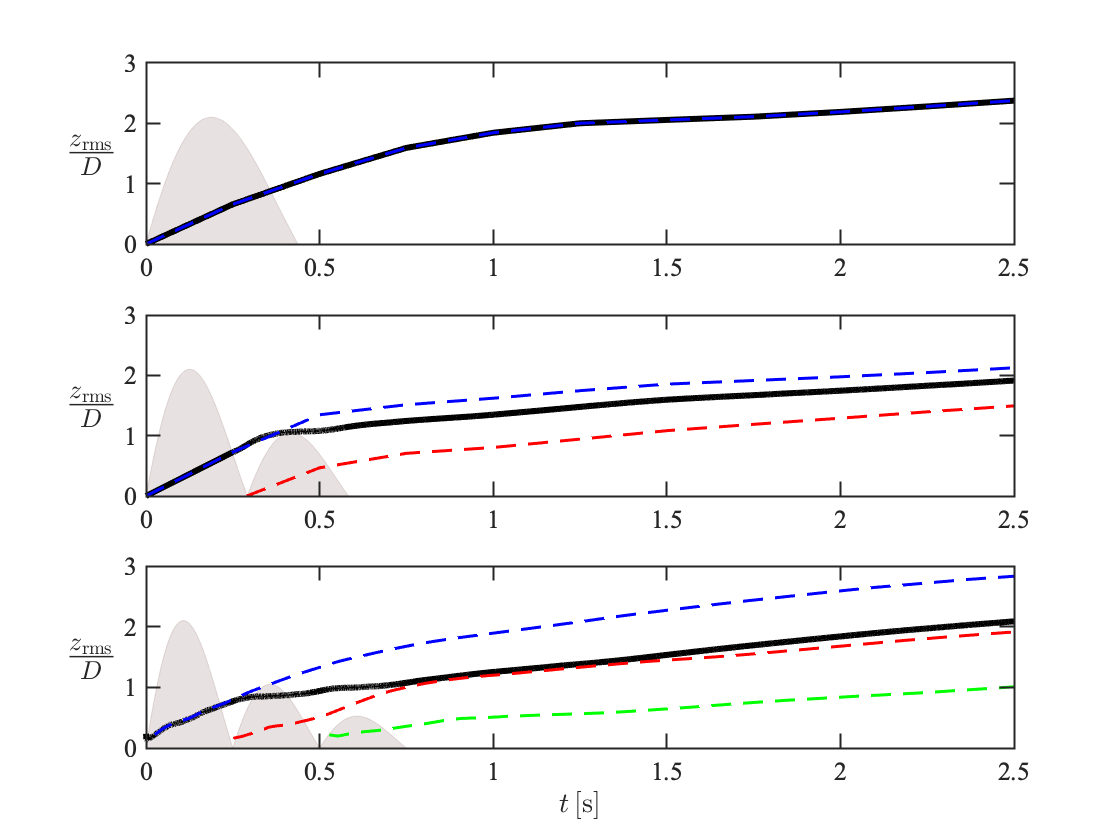}}
\caption{Temporal evolution of lateral dispersion within the entire cloud (\textcolor{black}{\bm{$-$}}) and first (\textcolor{blue}{\bm{$--$}}), second (\textcolor{red}{\bm{$--$}}), and third pulse (\textcolor{green}{\bm{$--$}})  for the one pulse (top), two pulse (middle), and three-pulse (bottom) cases. The flow rate $Q(t)$ associated with each pulse is shown in gray.}
\label{fig:z_penetration}
\end{figure}

Particle dispersion is characterized herein by the root-mean-square (rms) of lateral particle position, given by $z_{\rm rms}=\sqrt{\sum_{i=1}^{N_p}{z_p^{(i)}}^2/N_p}$ . The temporal evolution of the cloud penetration, $x_c$, and dispersion, $z_{\rm rms}$, associated with each pulse are shown in Figs.~\ref{fig:x_penetration} and~\ref{fig:z_penetration}.  During the early stage injection, $x_c$ and $z_{\rm rms}$ are seen to oscillate in the two- and three-pulse cases. Additionally, the final displacement of $x_c$ and $z_{\rm rms}$ at $t=2$ s is lower than that of the single-pulse case. 
These observations can be attributed to the decreasing volumetric flow rate between each pulse, and as a consequence the average injection velocity is lower compared to the single-pulse case. After the pulsatile injection completes, however, the streamwise displacement of the cloud associated with later pulses is seen to `catch up' with the earlier pulses despite being injected at later time and with significantly lower velocity. For example, the geometric center of the second-pulse particles nearly coincide with the overall particle cloud for the three-pulse case at $t = 2$ s, indicating once again that the penetration of potentially more contagious particles from later pulses are accelerated by earlier pulses. On the contrary, such effect of pulsatility is not observed for the lateral dispersion for which particles from earlier pulses disperse further.

\begin{figure}[!h]
\captionsetup[subfigure]{labelformat=simple}
\centering
{\includegraphics[width=.85\textwidth]{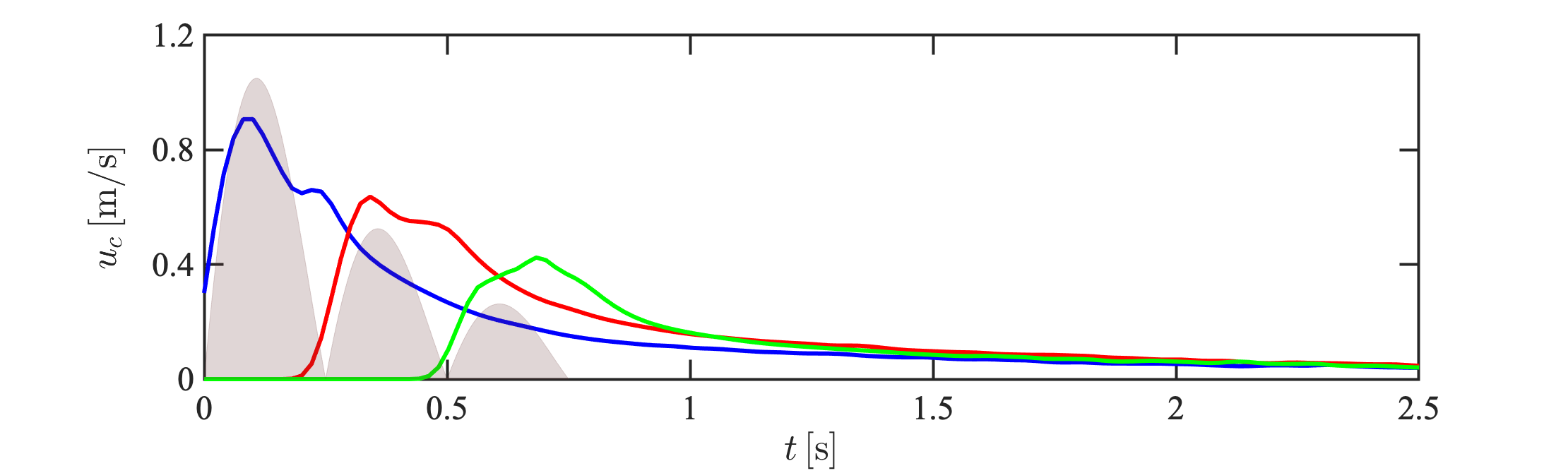}}
\caption{Streamwise velocity of the geometric center of the 
first (\textcolor{blue}{\bm{$-$}}), second (\textcolor{red}{\bm{$-$}}), and third pulse (\textcolor{green}{\bm{$-$}}) of a three-pulse cough. The flow rate $Q(t)$ associated with each pulse is shown in gray.}
\label{fig:vel_center}
\end{figure}

The velocity of the geometry center of each pulse associated with the three-pulse case, $u_c={\rm d}x_c/{\rm d}t$, is shown in Fig.~\ref{fig:vel_center}. Due to the finite inertia of the particles, $u_c$ lags the fluid velocity during early stage injection. It can be seen that the peak velocity of the third pulse is only reduced by a factor of two compared to the first pulse, despite its injection velocity being reduced by a factor of four. In addition, the velocity associated with the second pulse exceeds the velocity of the first pulse when $t>0.3$ s, further demonstrating the ability of particles injected at later times to catch up to particles near the front of the cloud. Further downstream of the inlet, the velocities of each geometric center converge, consistent with observations from human subjects~\citep{abkarian2020speech} that the unsteadiness of expiratory events diminishes far from the mouth.

\begin{figure}[!h]
\captionsetup[subfigure]{labelformat=simple}
\centering
{
    \includegraphics[width=.33\textwidth]{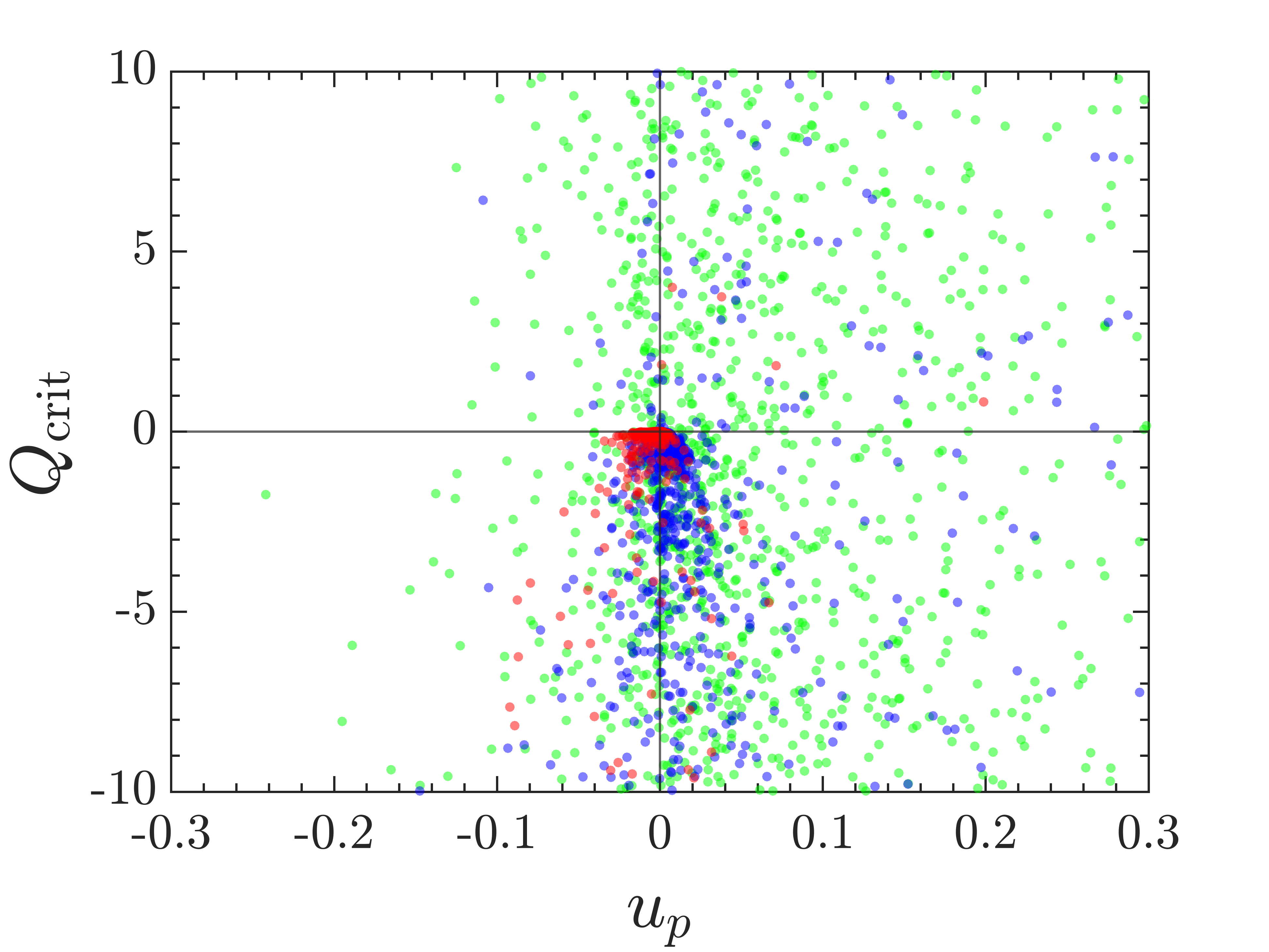}\hfill
    \includegraphics[width=.33\textwidth]{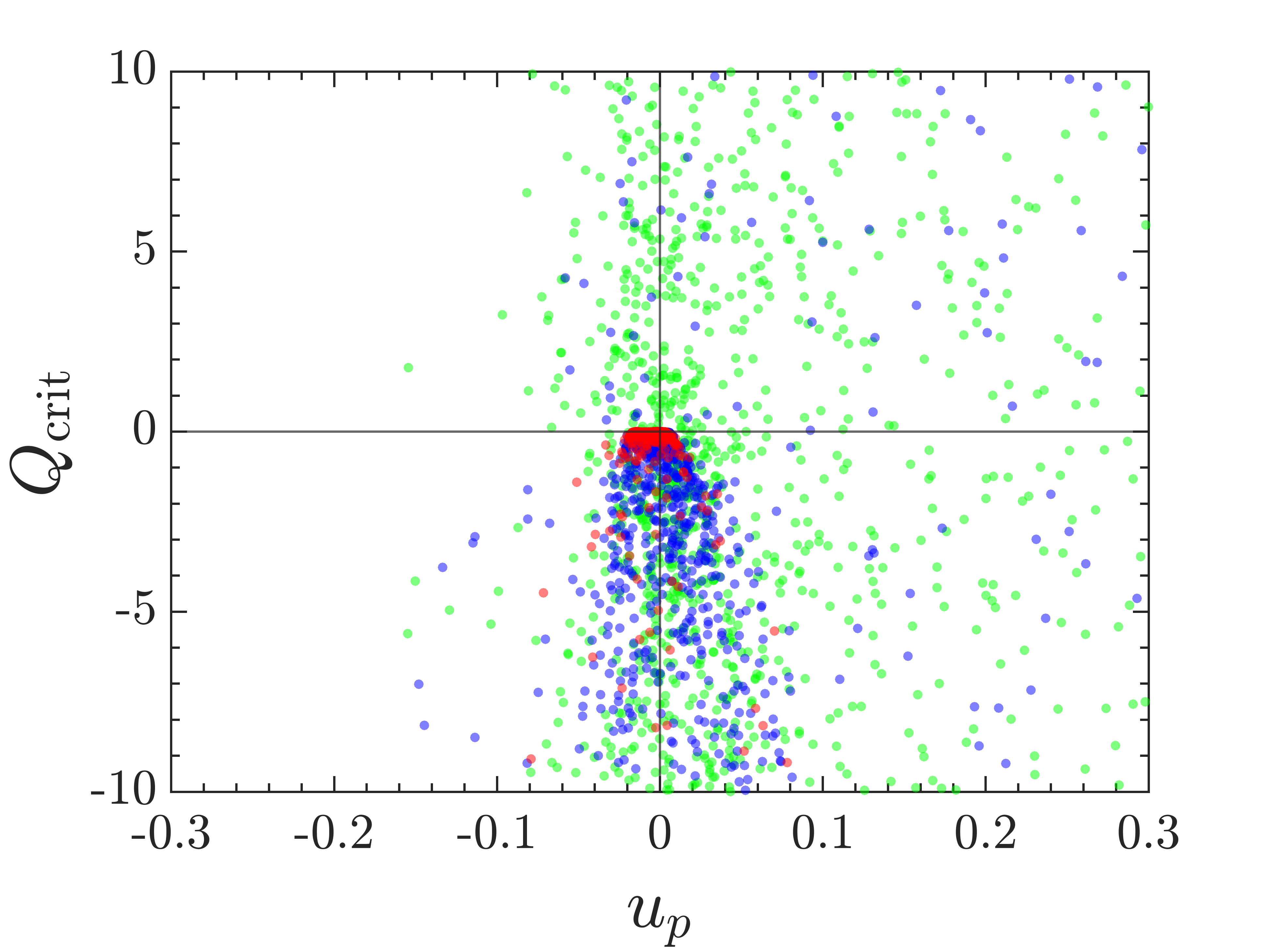}\hfill
    \includegraphics[width=.33\textwidth]{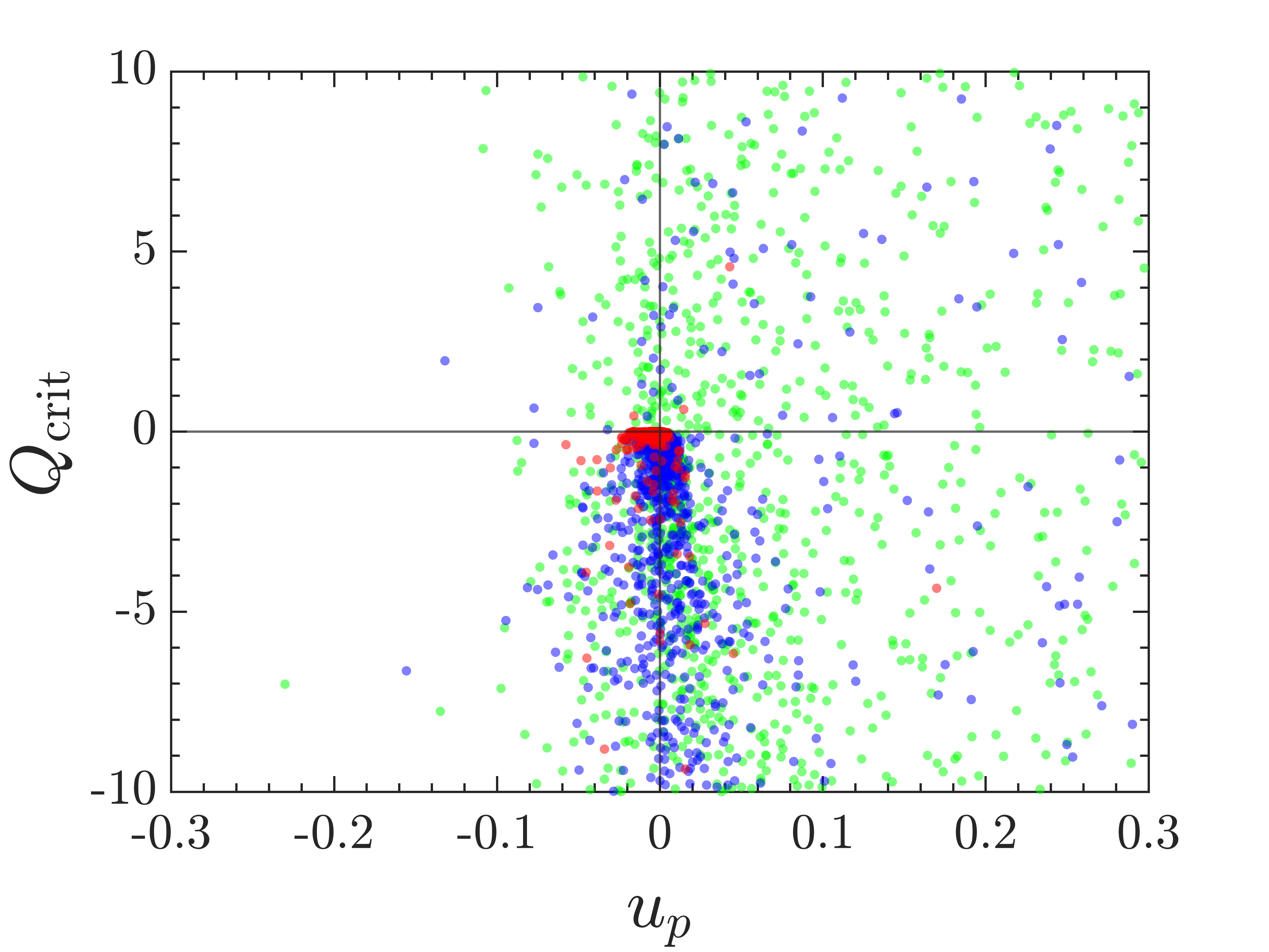}\hfill
    \includegraphics[width=.33\textwidth]{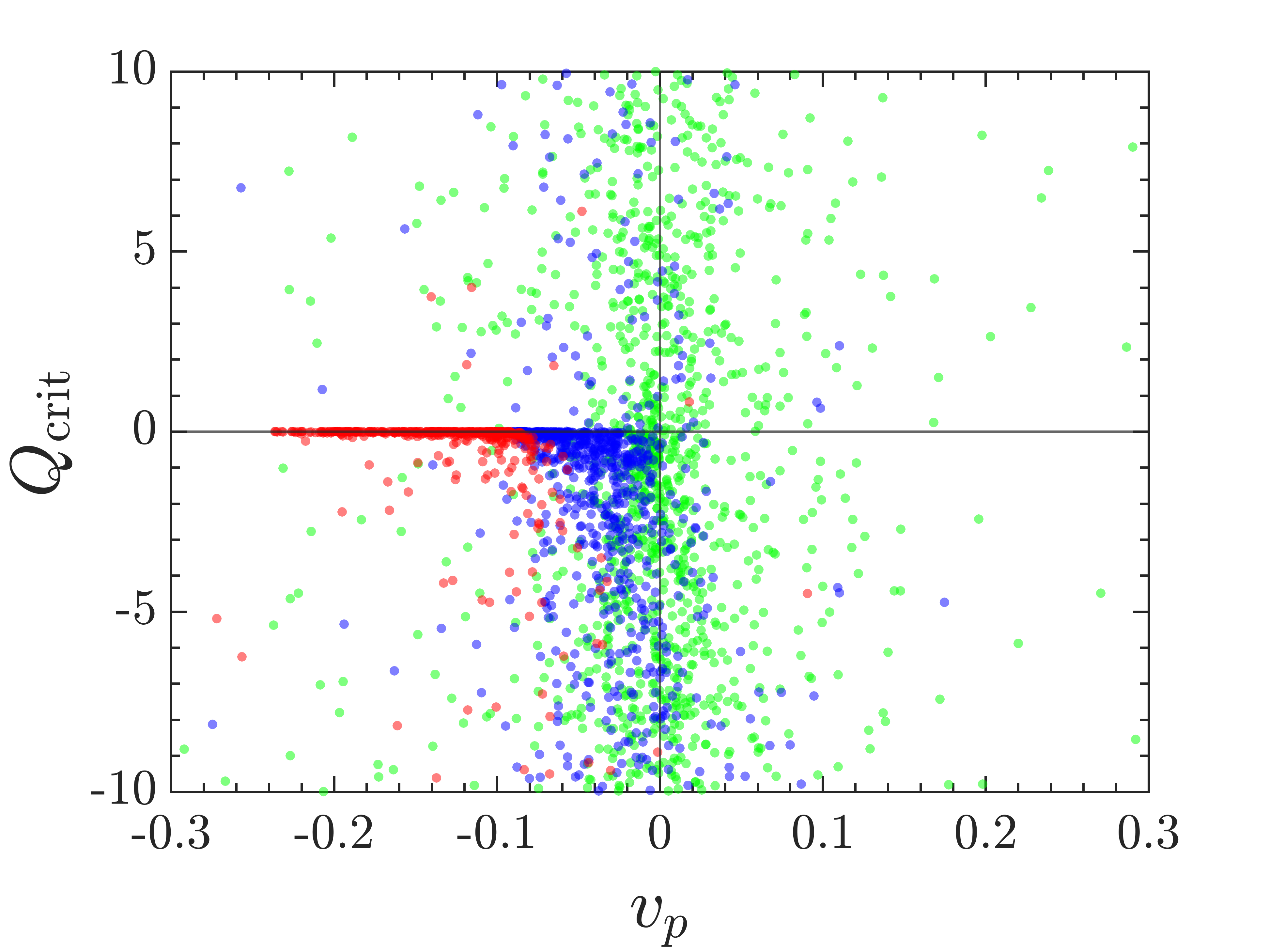}\hfill
    \includegraphics[width=.33\textwidth]{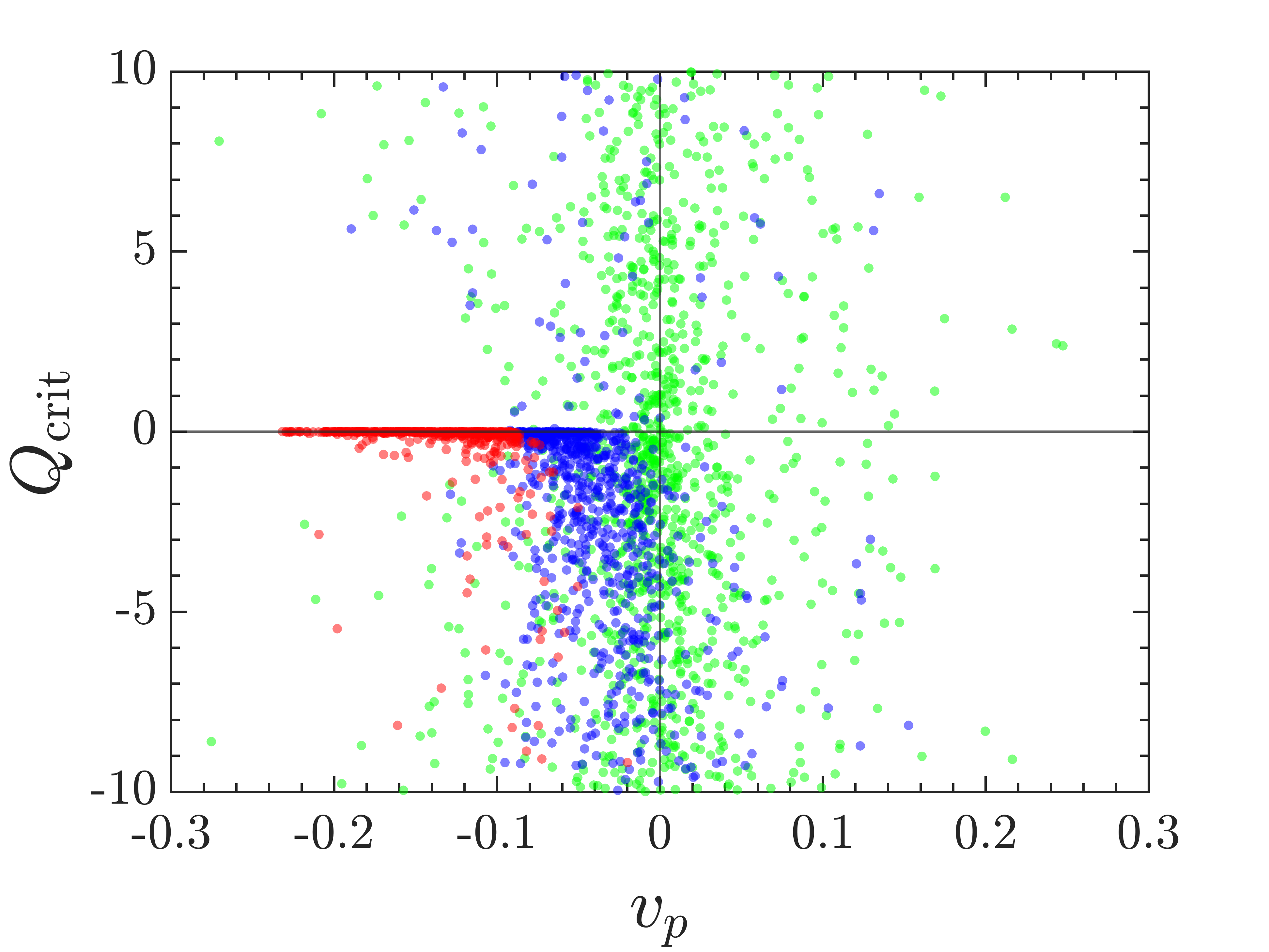}\hfill
    \includegraphics[width=.33\textwidth]{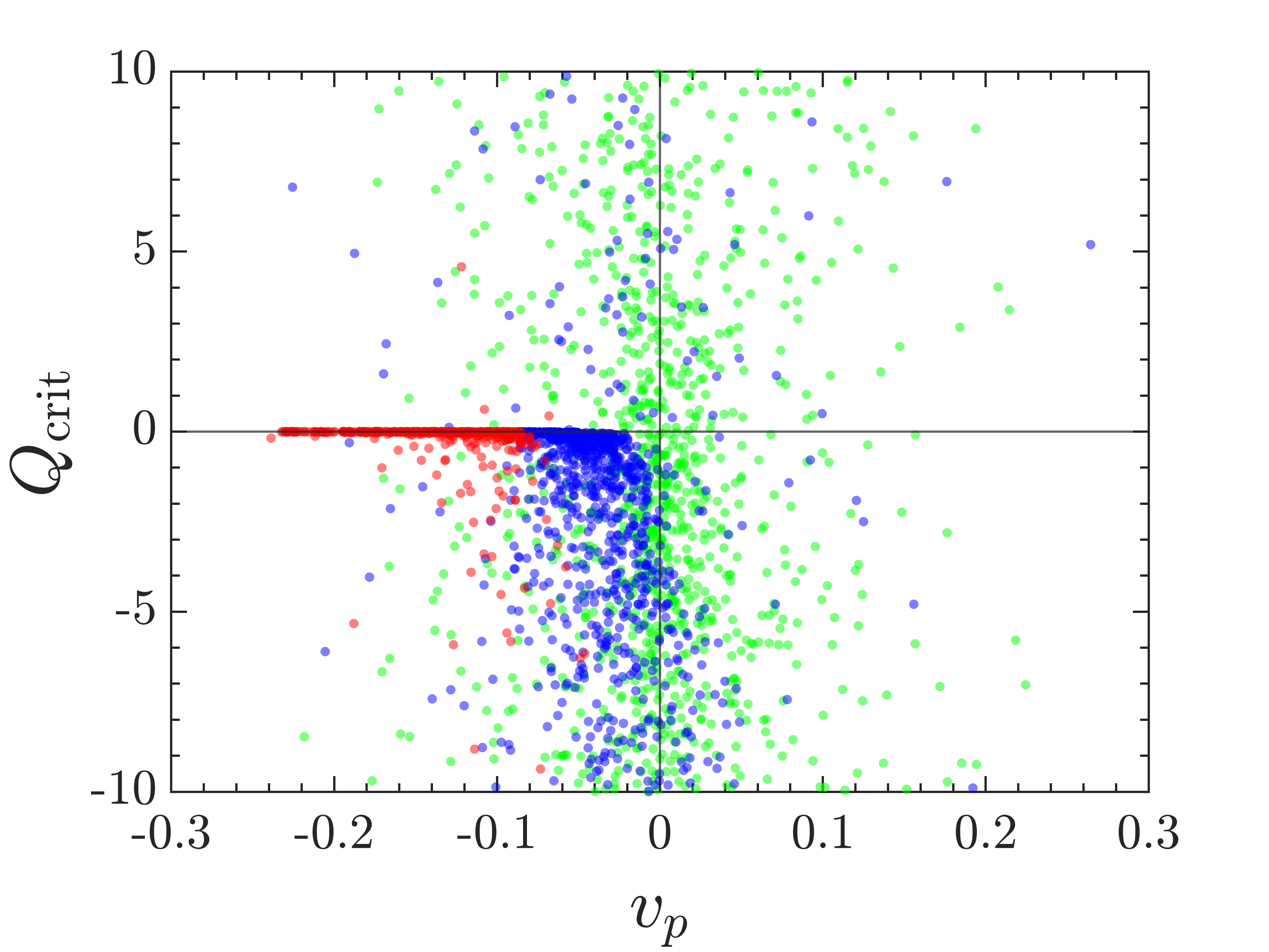}\hfill
    \includegraphics[width=.33\textwidth]{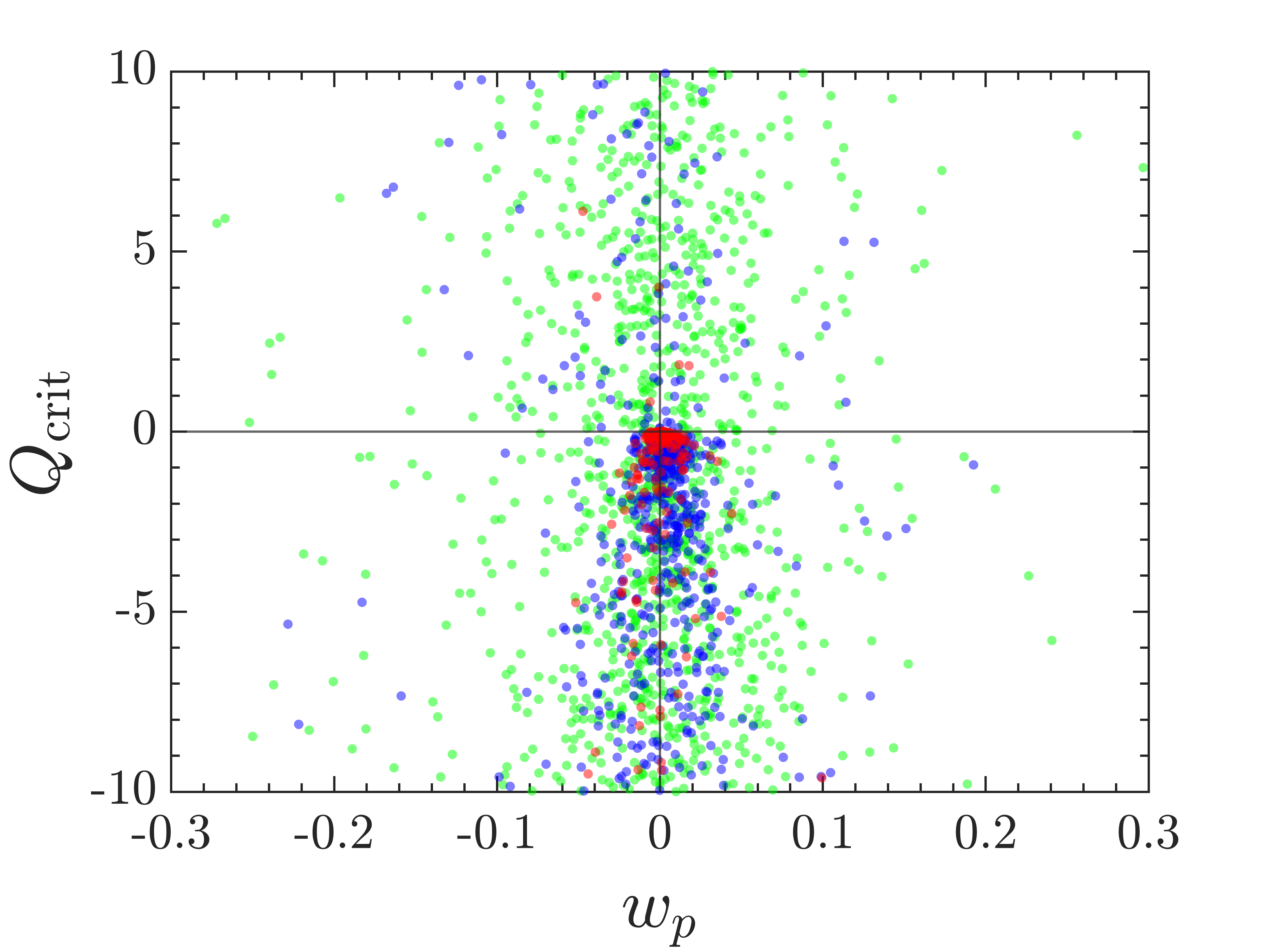}\hfill
    \includegraphics[width=.33\textwidth]{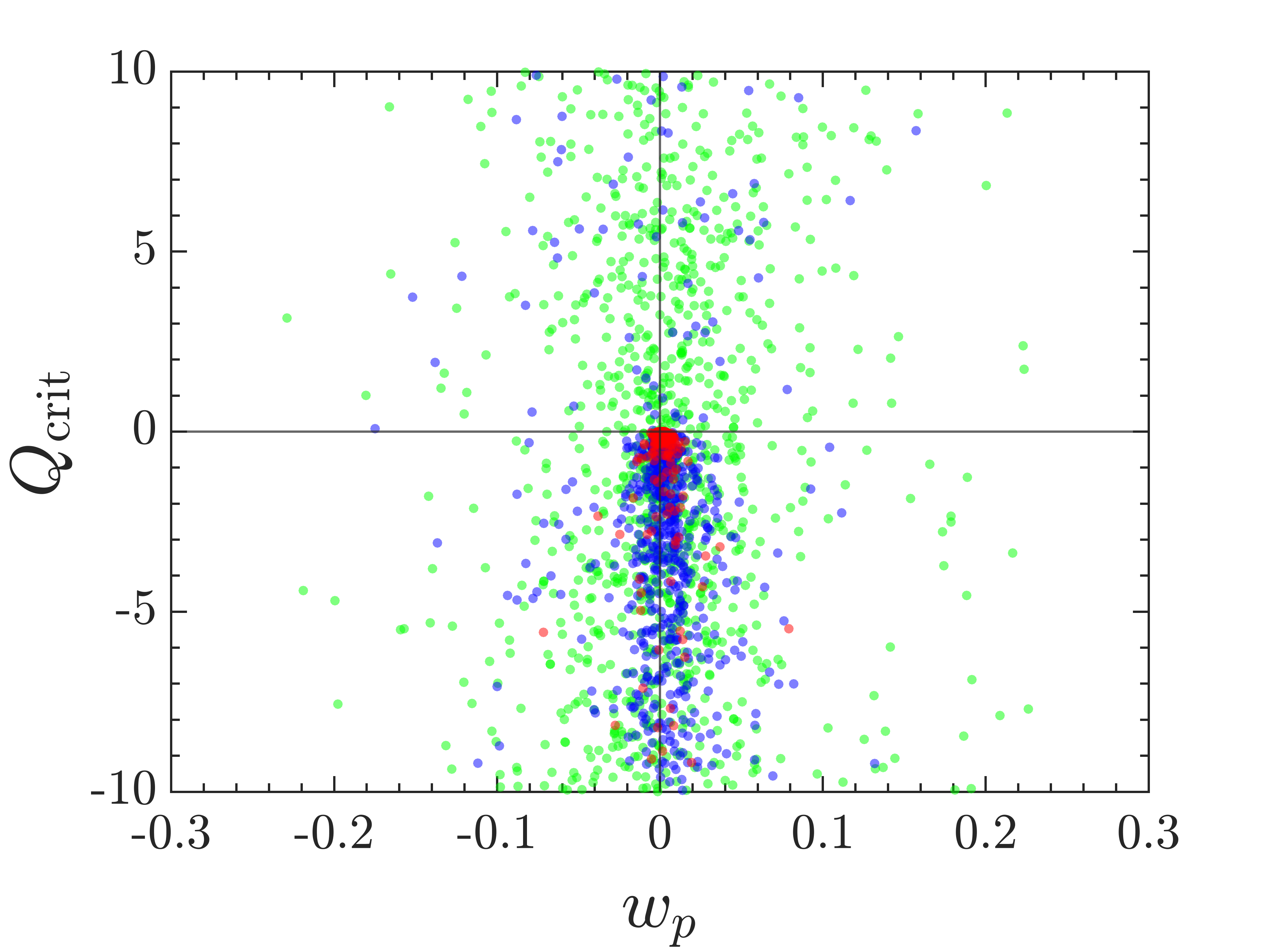}\hfill
    \includegraphics[width=.33\textwidth]{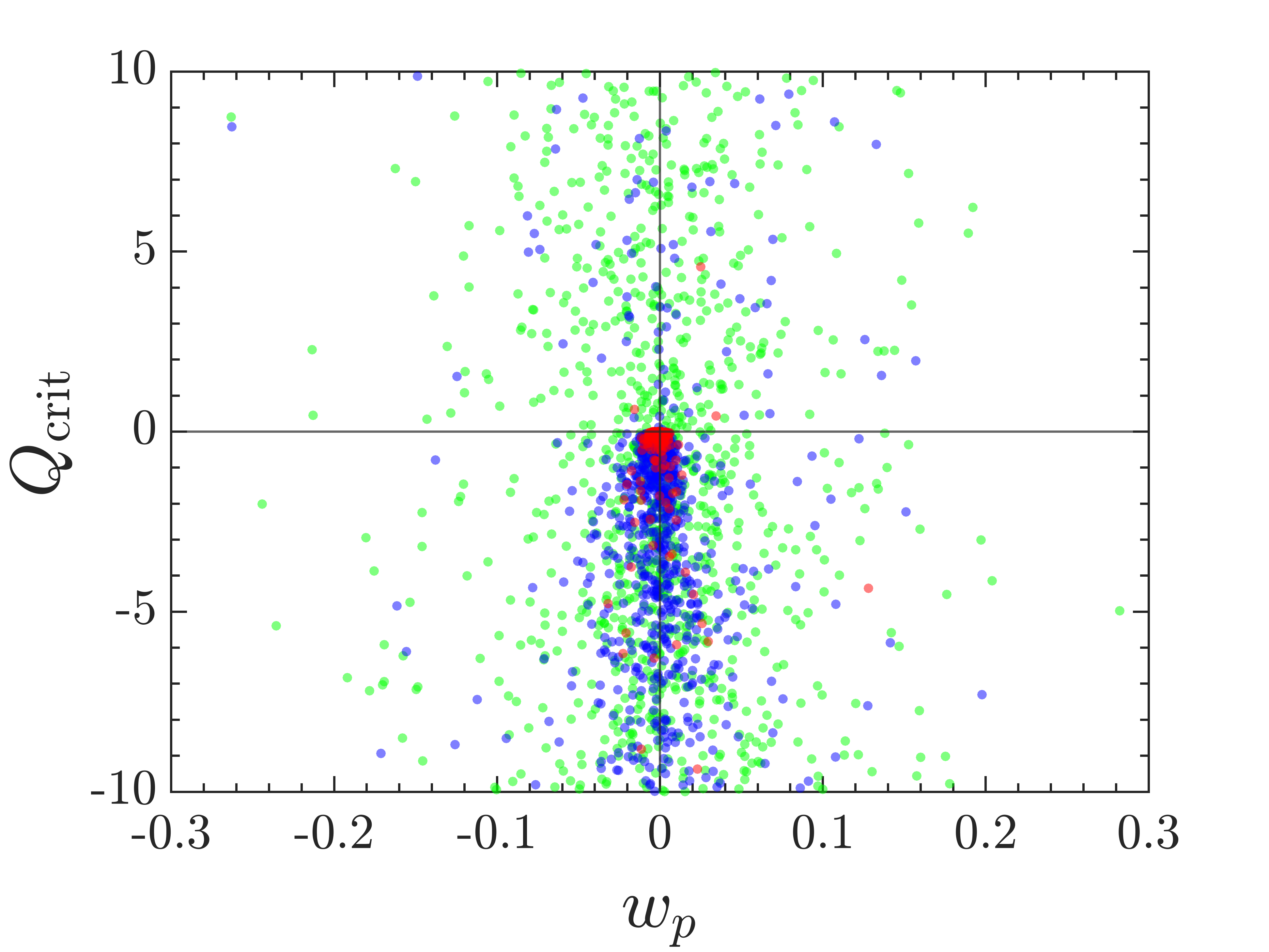}
    \\[\smallskipamount]
}
\caption{Scatter plots of the $Q$-criterion against directional velocities in the $x$ (top row), $y$ (middle row), and $z$ direction (bottom row), at $t=1$ s. Left, middle, and right columns correspond to the one pulse, two pulse, and three pulse simulations. Particles are colored by size with $d_p \in [1, 30]$ $\upmu$m (green), $[30, 60]$ $\upmu$m (blue), and $[60, 100]~\upmu$m (red).}
\label{fig:scatter_plot1s}
\end{figure}

In studying the entrainment characteristics of particle-laden channel flows, \citet{marchioli2009preference} presented a framework highlighting the use of instantaneous joint correlations of non-vanishing components of the fluctuating velocity gradient tensor to determine locations of particle preferential concentration. In addition, the relationship between particle size and their corresponding fluid topology was exploited to identify particle preferential sampling. The present work extends these techniques to correlate the range of particle sizes seen in a pulsatile cough with their immediate fluid environment and vortex structures. This is accomplished by correlating the value of the $Q$-criterion against a particle's directional velocity ($u_p, v_p, w_p$) (see Fig.~\ref{fig:scatter_plot1s}). The sign of $Q_{\rm crit}$ describes the fluid environment that the particle currently resides, with $Q_{\rm crit} > 0$ corresponding to regions of high vorticity, $Q_{\rm crit} \approx 0$ corresponding to regions of no flow or constant strain, and $Q_{\rm crit} < 0$ corresponding to regions of high strain rate. The particle directional velocity describes the dispersive characteristics of the particles, with a large spread in velocity being associating with high directional dispersion. 

Particles are grouped into three size ranges as depicted in Fig.~\ref{fig:part_distro}. It is first observed that small particles ($d_p \in [1, 30]$ $\upmu$m) equally sample all values of $Q_{\rm crit}$, exhibiting no preferential location in terms of fluid vortical structures. Mid-sized particles ($d_p \in [30, 60]$ $\upmu$m) tend to sample regions of high strain rate, indicating their ejection from vorticity-dominated regions, i.e., classical preferential concentration. Large particles ($d_p \in [60, 100]$ $\upmu$m) are seen to sample regions of constant strain ($Q_{\rm crit} = 0$) as a consequence of them falling out of the cloud due to gravity. Note that the distribution of mid-sized particles are skewed to negative values of $v_p$, while this is not observed for small particles, indicating that gravity has an effect on the former but not the latter. 

It can also be seen that the distribution in lateral velocity, $w_p$, associated with mid-sized particles is narrower in the cases with multiple pulses compared to the single-pulse case. Specifically, mid-sized particles are preferentially sampling near-zero values of $w_p$ for a wider range of $Q_{\rm crit}$ compared to the one-pulse case. The distribution in the gravity-aligned ($y$) direction remains unchanged between the three cases, which is expected as gravity plays a more important role for mid- and large-sized particles in this direction. For the scatter plots in $x$-direction, although most particles in all three cases are preferentially sampling the right half plane ($u_p>0$) as the net particle flux is positive in the streamwise direction, more mid-sized particles are seen to have negative $u_p$ over a wider range of negative $Q_{\rm crit}$. These aforementioned differences in $x$ and $z$ are likely a result of mid-sized particle being entrained by the vortices generated by the subsequent pulses as they respond most effectively to turbulent eddies due to their intermediate Stokes numbers. Consequently, it also explains the decreasing penetration ($x_c$) and dispersion ($z_{\rm rms}$) with increasing pulsatility, as seen in Figs.~\ref{fig:x_penetration} and \ref{fig:z_penetration}. 

\subsection{Theoretical modeling of respiratory emissions}\label{sec:cloud-models}

\citet{bourouiba2014violent} proposed a theoretical model for the cloud penetration based on the notion of conservation of cloud momentum. It has been generally observed from their experiments that two phases of the cloud evolution exist. The first phase is dominated by jet-like dynamics, corresponding to the high-speed release of the fluid-particle mixture. In this phase, penetration is modeled as a function of specific momentum flux $M$ according to
\begin{equation}\label{eq:jet-regime}
\frac{\mathrm{d} x_c}{\mathrm{d} t}=C_{M} \sqrt{\frac{M}{{\alpha_{1}^{'}}^{2} x_c^{2}}} {\quad \rm or\quad } x_c(t) = \left( \frac{4 C_M^2 M}{{\alpha_{1}^{'}}^{2} }\right)^{1/4} t^{1/2},
\end{equation}
where $C_M$ is a constant coefficient of the first regime. The second phase is dominated by ``puff-like'' dynamics, characterized by the self-similar growth of the puff cloud. The puff penetration evolution is given by
\begin{equation}\label{eq:puff-regime}
\frac{\mathrm{d} x_c}{\mathrm{d} t}=C_{I} \frac{I}{{\alpha_{2}^{'}}^{2}  x_c^{3}} {\quad \rm or\quad } x_c(t) = \left( \frac{4 C_I I}{{\alpha_{2}^{'}}^{3} }\right)^{1/4} t^{1/4},
\end{equation}
where $C_I$ is a constant coefficient of the second regime, and the total specific momentum of the cloud, $I$, is defined as
\begin{equation}
I(t)= \int_0^t M\,{\rm d}\tau.
\end{equation}
Here, $\alpha_1^{'}$ and $\alpha_2^{'}$ are particle entrainment coefficients, which satisfy $r_c = \alpha^{'}x_c$ where $r_c = \sum_{i=1}^{N_p}\sqrt{{y_p^{(i)}}^2+{z_p^{(i)}}^2}/N_p$ is the geometric mean radius of the particle cloud.

\begin{figure}[h!]
\begin{center}
\begin{subfigure}[t]{0.329\textwidth}
\includegraphics[width=\textwidth]{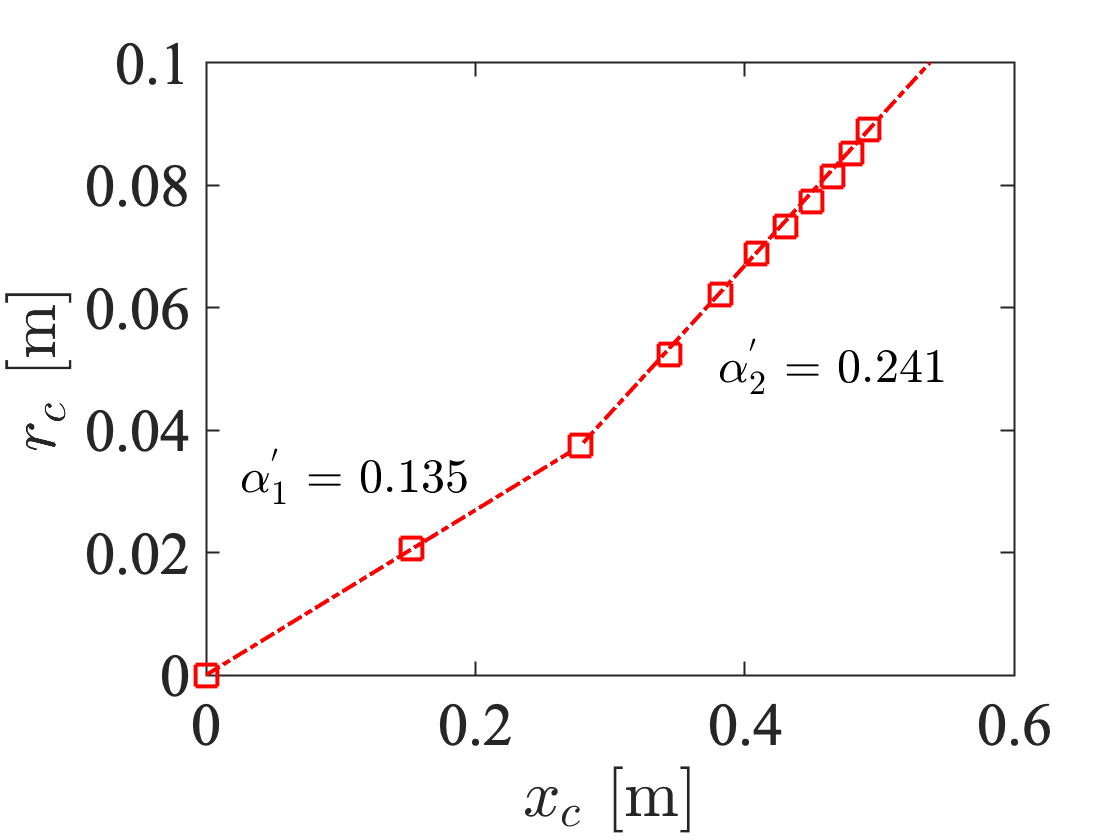}
\caption{One-pulse}
\label{fig:alpha_1p_all}
\end{subfigure}
\begin{subfigure}[t]{0.329\textwidth}
\includegraphics[width=\textwidth]{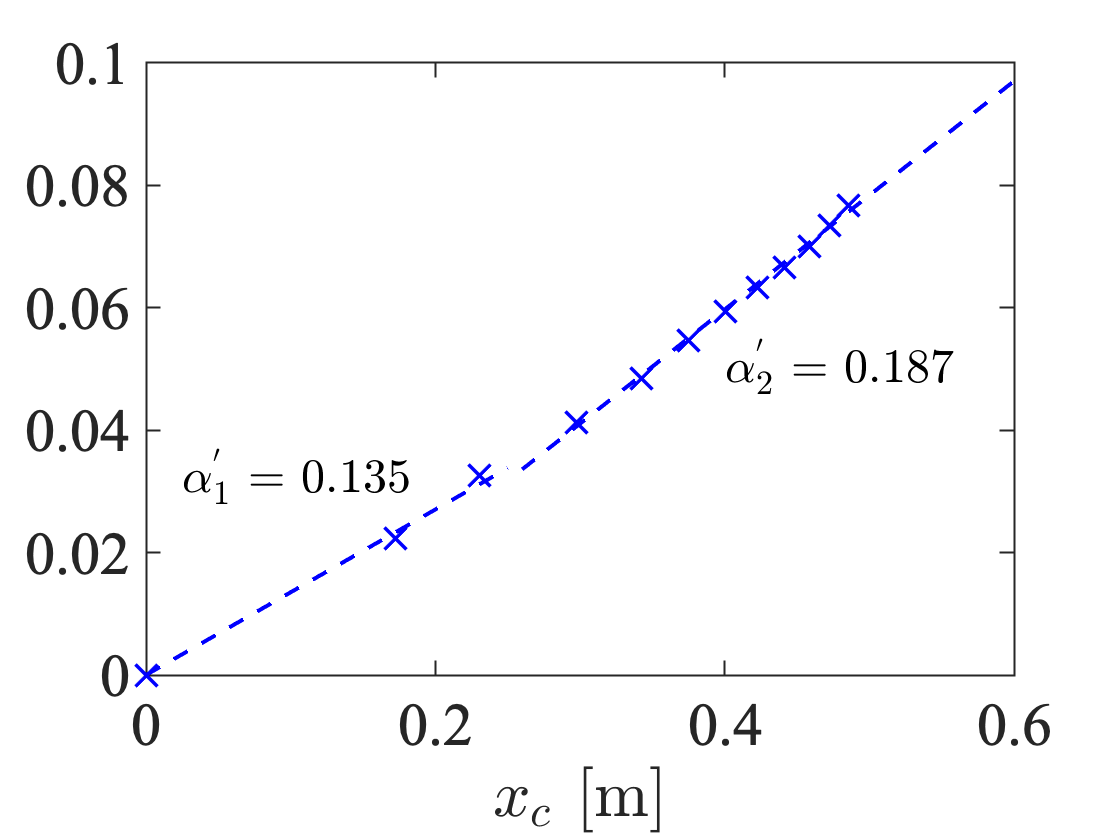}
\caption{Two-pulse}
\label{fig:alpha_2p_all}
\end{subfigure}
\begin{subfigure}[t]{0.329\textwidth}
\includegraphics[width=\textwidth]{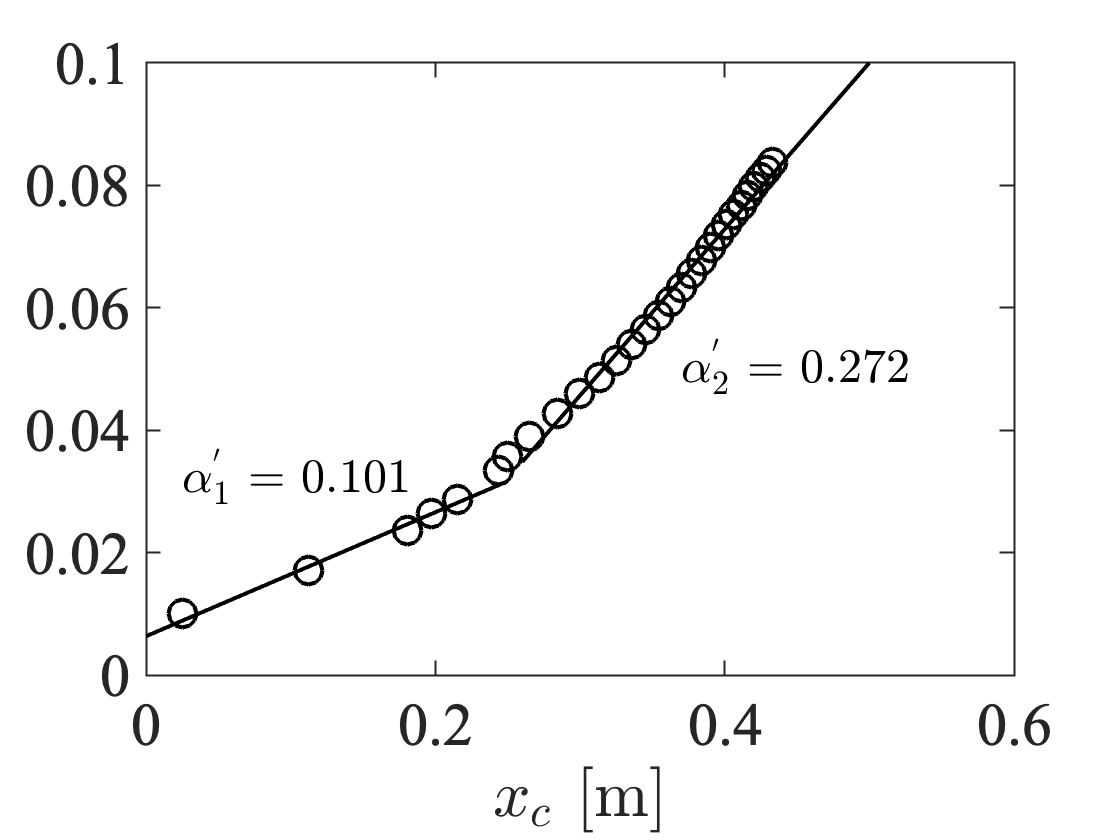}
\caption{Three-pulse}
\label{fig:alpha_3p_all}
\end{subfigure}
\begin{subfigure}[t]{0.329\textwidth}
\includegraphics[width=\textwidth]{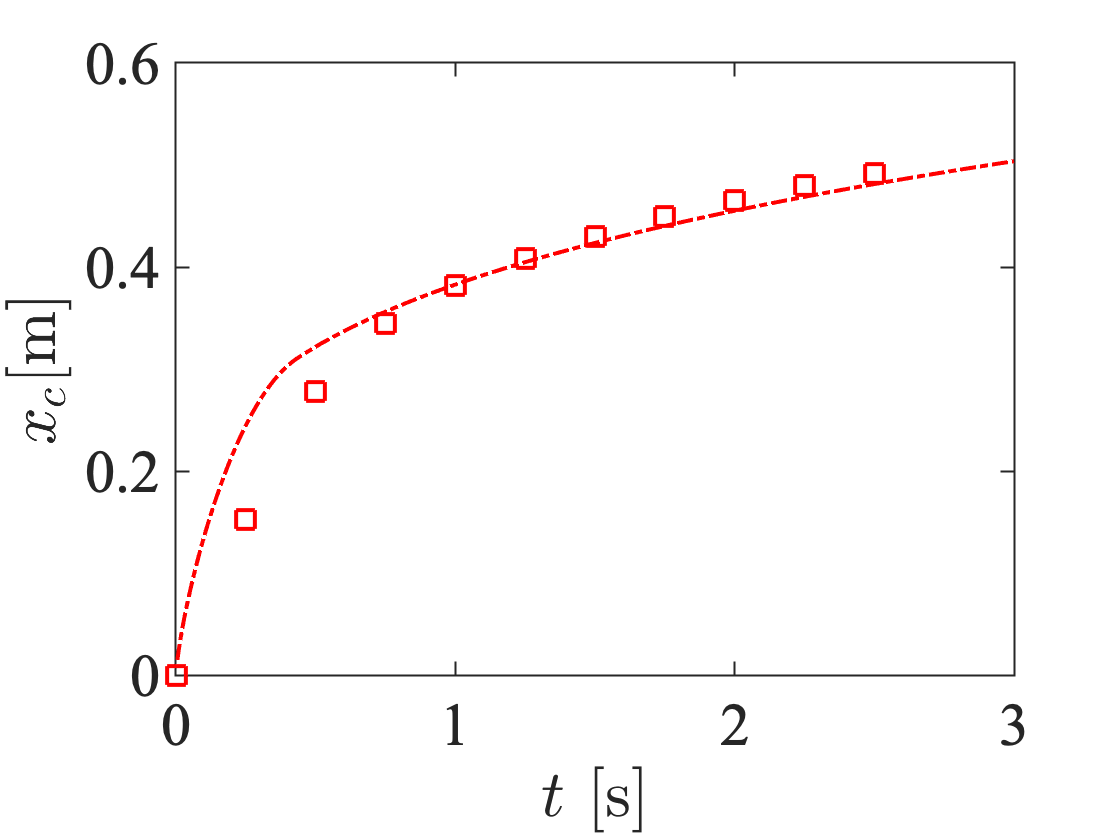}
\caption{One-pulse}
\label{fig:xcom_1p_all}
\end{subfigure}
\begin{subfigure}[t]{0.329\textwidth}
\includegraphics[width=\textwidth]{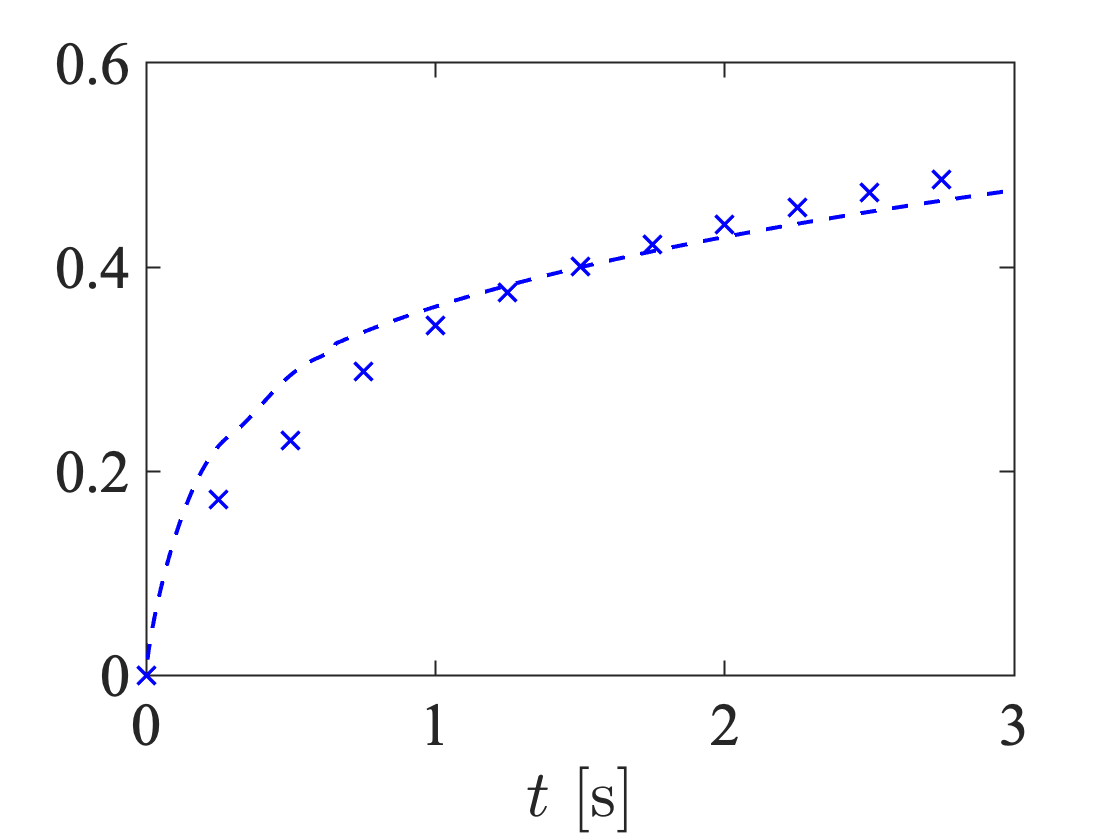}
\caption{Two-pulse}
\label{fig:xcom_2p_all}
\end{subfigure}
\begin{subfigure}[t]{0.329\textwidth}
\includegraphics[width=\textwidth]{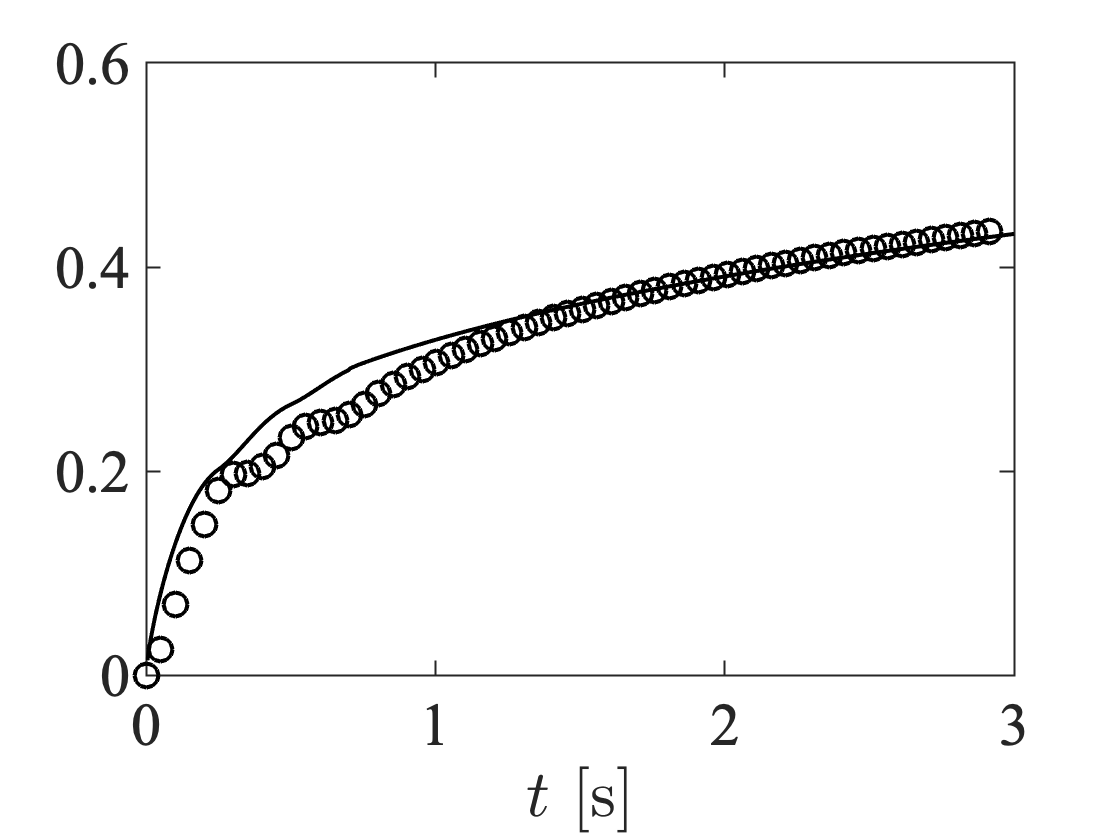}
\caption{Three-pulse}
\label{fig:xcom_3p_all}
\end{subfigure}
\caption{Top: mean radius of the cloud versus its geometric center. Bottom: geometric center of the cloud versus time. The simulation results are shown as symbols, linear fits (top) and model predictions (bottom) are shown as lines. Line types and colors same as Fig.~\ref{fig:pulse_profile}.}
\label{fig:alpha_fitting_all}
\end{center}
\end{figure}

The particle entrainment coefficients of the two regimes for all three cases are determined by the slope of $r_c$ versus $x_c$ as shown in Figs.~\ref{fig:alpha_1p_all}--\ref{fig:alpha_3p_all}. The entrainment coefficient of the second regime, $\alpha_2^{'}$, is observed to be larger and more sensitive to pulsatility than the entrainment coefficient of the first regime, $\alpha_1^{'}$, indicating that the vortex interactions may be more significant in the puff-like regime. Penetration can be predicted by combining Eqs.~\eqref{eq:jet-regime} and~\eqref{eq:puff-regime} using these values of $\alpha_1^{'}$ and $\alpha_2^{'}$. In contrast to \citet{bourouiba2014violent}, here $M$ is time-varying due to the pulsatile nature of the expiratory flow. To this end, the average specific momentum $\overline{M}$ is used in Eq.~\eqref{eq:jet-regime}, given by 
\begin{equation}
\overline{M}(t) =  \frac{1}{t}\int_0^t M\,{\rm d}\tau,
\end{equation}
where $M$ is assumed to follow the pulsatile profile given by Eq.~\eqref{damped_pulse}, i.e., ${M}(t) = M_0 \lvert e^{-t/ \tau} \sin(\omega t) \rvert$. The coefficients $C_I$ and $C_M$ are determined by least-square fitting of the puff regime and solving $( 4 C_{M}^2 M/{\alpha_{1}^{'}}^{2} ) ^{1/4} t_{\rm cr}^{1/2}=( 4C_{I} I/{{\alpha_{2}^{'}}^{3}} ) t_{\rm cr}^{1/4}$ respectively, with $t_{\rm cr}$ the intersection time of the two regimes.

The model predictions are displayed in Figs.~\ref{fig:xcom_1p_all}--\ref{fig:xcom_3p_all}. It can be seen that the second puff-like regime for all three cases scale as $x_c \sim t^{1/4}$. For the first jet-like regime, however, the penetration profiles deviate from $x_c \sim t^{1/2}$ and instead exhibit oscillations for the pulsatile cases, which is not correctly captured by the model.

\begin{figure}[h!]
\begin{center}
\begin{subfigure}[t]{0.329\textwidth}
\includegraphics[width=\textwidth]{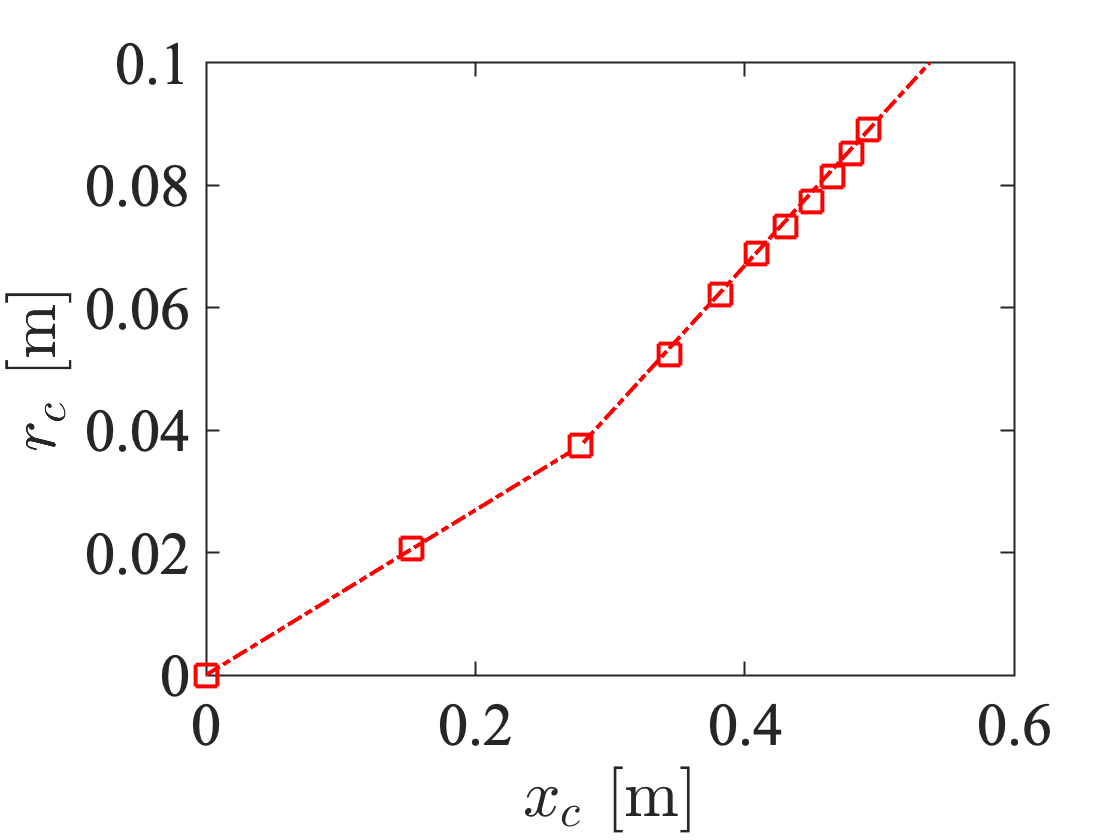}
\caption{One-pulse}
\label{fig:alpha_1p}
\end{subfigure}
\begin{subfigure}[t]{0.329\textwidth}
\includegraphics[width=\textwidth]{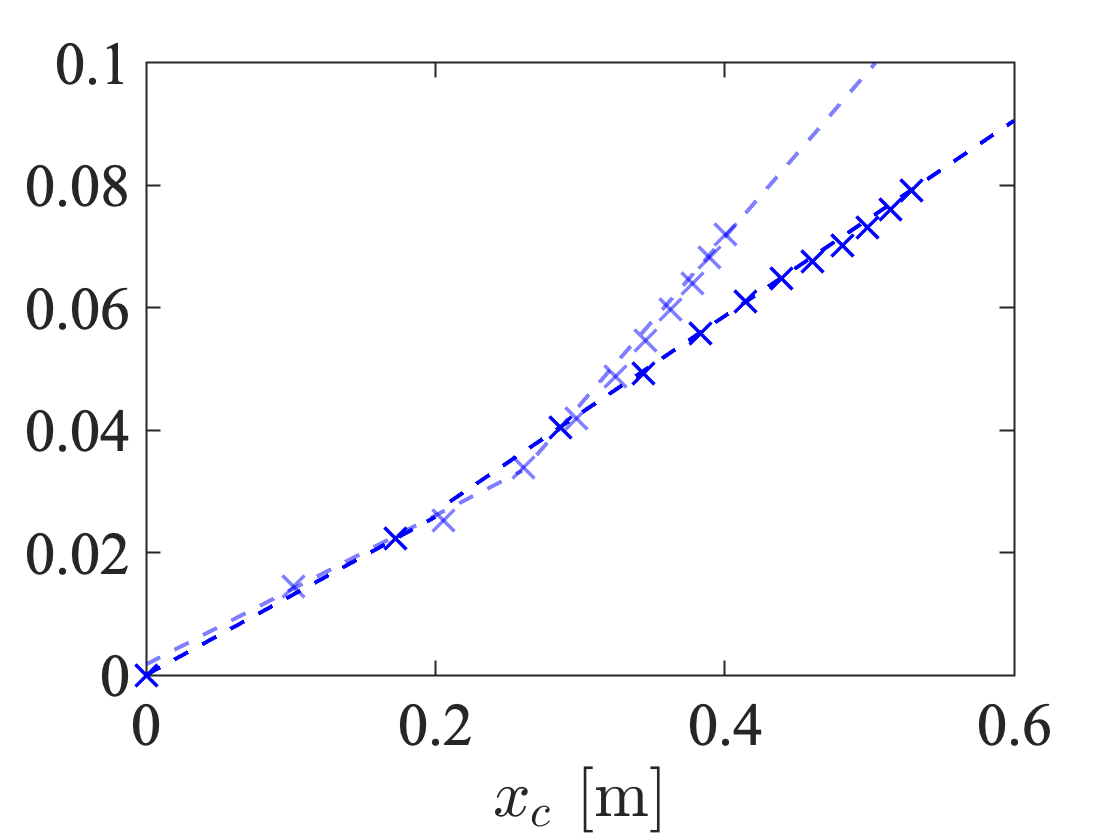}
\caption{Two-pulse}
\label{fig:alpha_2p}
\end{subfigure}
\begin{subfigure}[t]{0.329\textwidth}
\includegraphics[width=\textwidth]{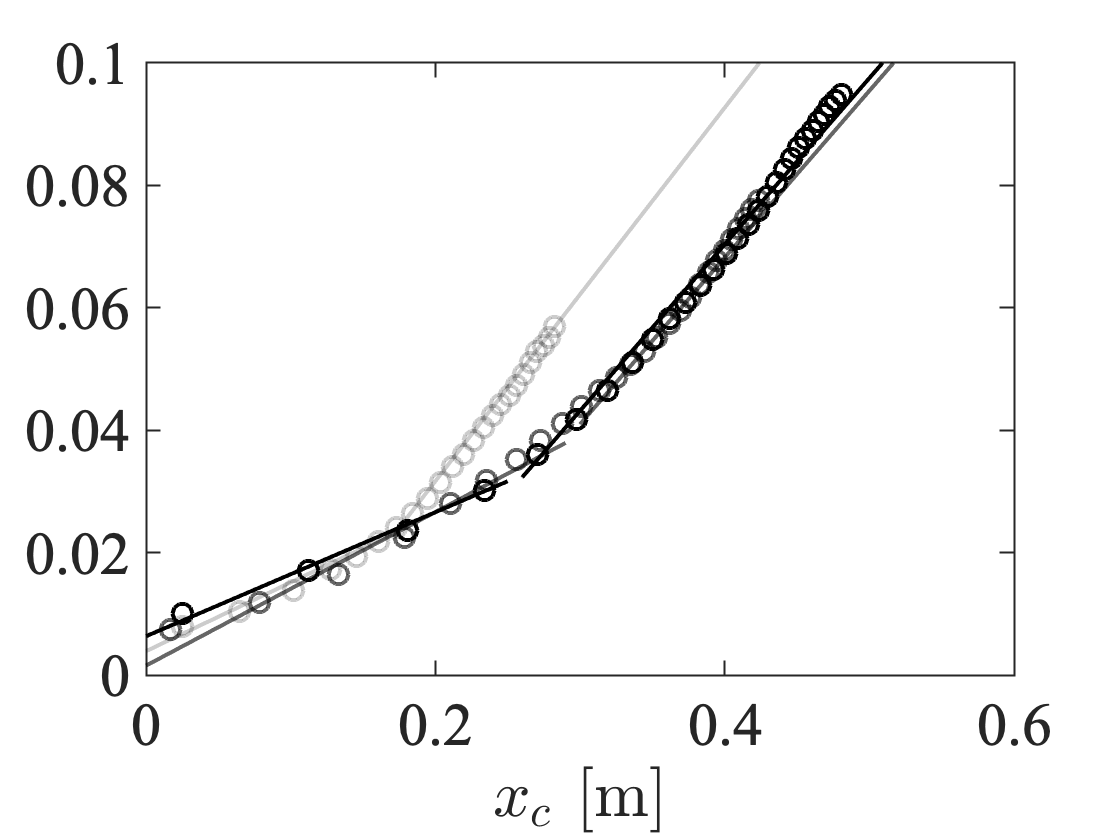}
\caption{Three-pulse}
\label{fig:alpha_3p}
\end{subfigure}
\begin{subfigure}[t]{0.329\textwidth}
\includegraphics[width=\textwidth]{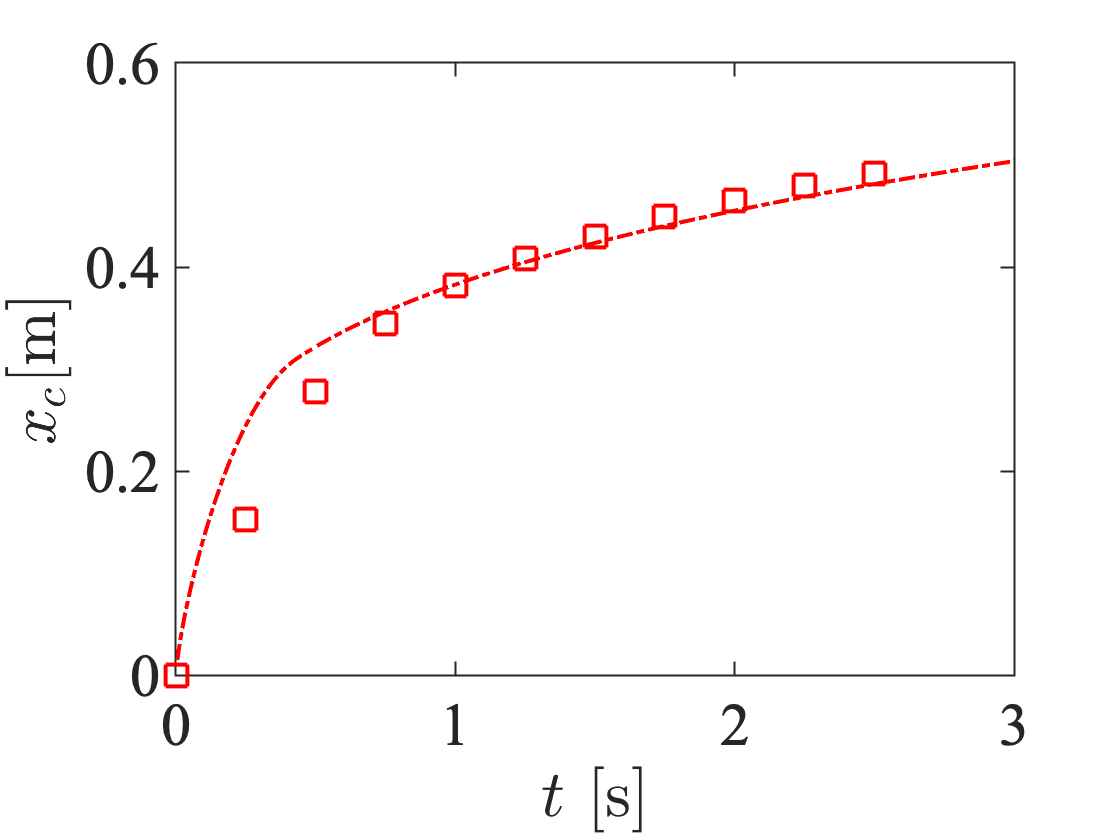}
\caption{One-pulse}
\label{fig:xcom_1p}
\end{subfigure}
\begin{subfigure}[t]{0.329\textwidth}
\includegraphics[width=\textwidth]{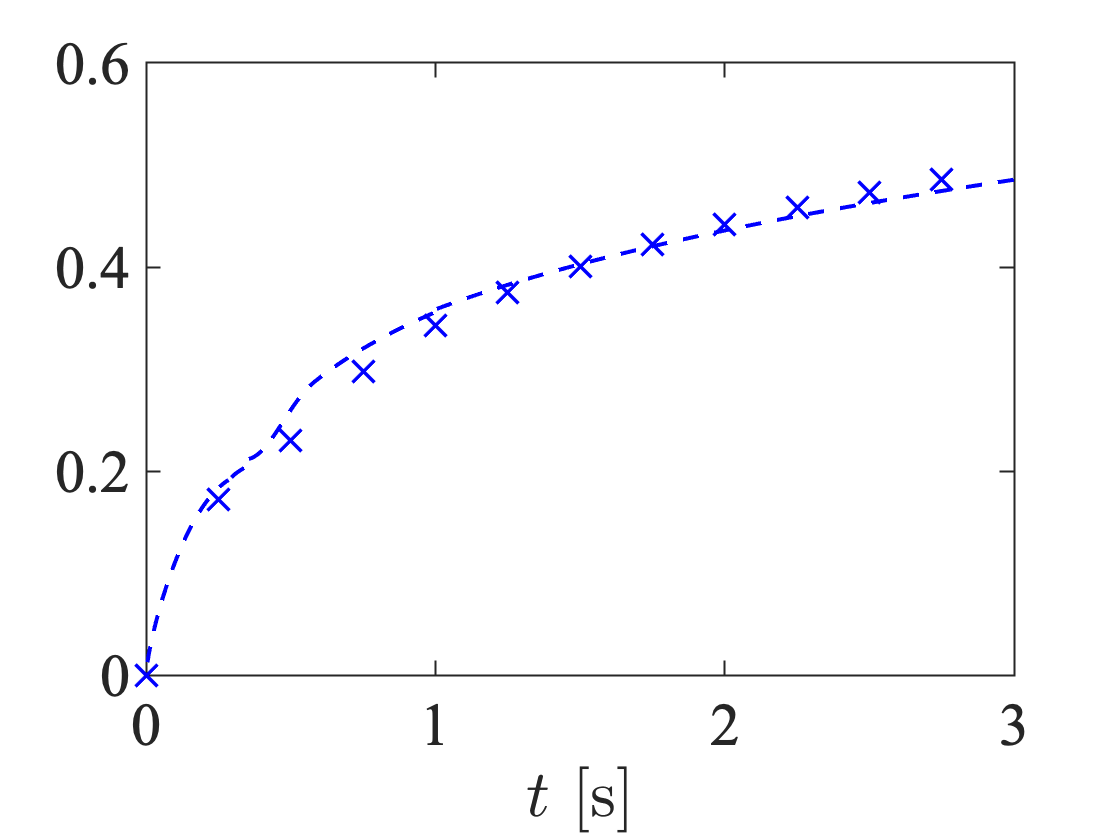}
\caption{Two-pulse}
\label{fig:xcom_2p}
\end{subfigure}
\begin{subfigure}[t]{0.329\textwidth}
\includegraphics[width=\textwidth]{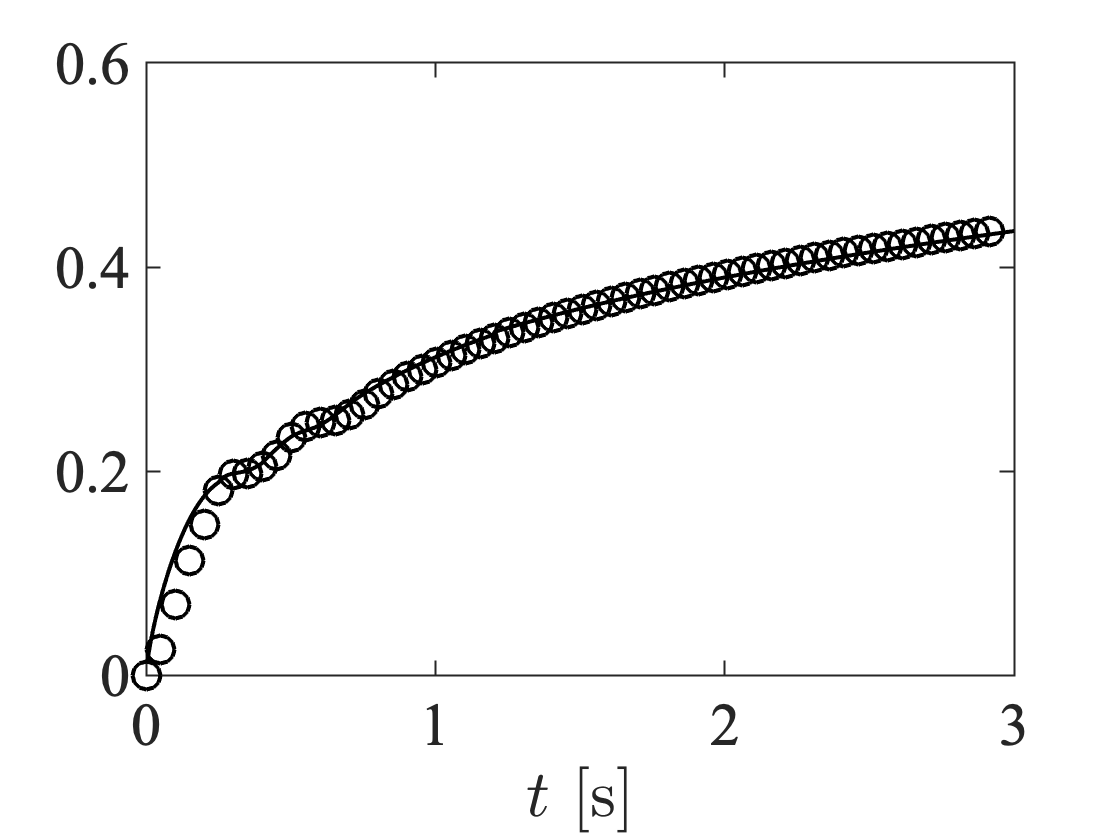}
\caption{Three-pulse}
\label{fig:xcom_3p}
\end{subfigure}
\caption{Top: mean radius of the cloud versus its geometric center. Bottom: geometric center of the cloud versus time. The simulation results are shown as symbols, linear fits (top) and model predictions (bottom) are shown as lines. Line types and colors same as Fig.~\ref{fig:pulse_profile}. Data from the second and third pulse are shown in increasing transparency.}
\label{fig:alpha_fitting}
\end{center}
\end{figure}

Here, we extend the current model to account for pulsatility. Instead of applying the conservation law to the entire cloud, the particle concentration coefficient of each pulse is extracted separately for the pulsatile cases. As shown in Figs.~\ref{fig:alpha_1p}--\ref{fig:alpha_3p}, the particle cloud from each pulse follows their own two-stage evolution. In addition, $\alpha_1^{'}$ is observed to be similar between different pulses, whereas $\alpha_2^{'}$ of the second and third pulse are significantly larger than the first pulse, indicating a larger dispersion for late-injected particles as facilitated by the earlier pulses despite the overall dispersion is hindered. Using these values of $\alpha_1^{'}$ and $\alpha_2^{'}$, the penetration of each pulse is then modeled by following the same procedure described earlier. Let $x_c^{1}$, $x_c^2$, and $x_c^3$ denote the penetration of the first, second and third pulse, and $t_{1}$, $t_{2}$, and $t_{3}$ denote the time when the first, second and third pulse complete, the penetration of the entire cloud for the two-pulse case is modeled as the weighted average of each pulse given by
\begin{equation}
x_c(t) = 
\begin{dcases}
x_c^1(t), & \quad {\rm if\ }t \le t_{1}\\
\frac{I(t_{1})}{I(t)} x_c^1(t) + \frac{I(t) -I(t_{1})}{I(t)} x_c^2(t), & \quad {\rm if\ }t>t_{1},
\end{dcases}
\end{equation}
and similarly for the three-pulse case
\begin{equation}
x_c(t) = 
\begin{dcases}
x_c^1(t), & \quad {\rm if\ }t \le t_{1}\\
\frac{I(t_{1})}{I(t)} x_c^1(t) + \frac{I(t) -I(t_{1})}{I(t)} x_c^2(t), & \quad {\rm if\ }t_{1} < t < t_{2}\\
\frac{I(t_{1})}{I(t)} x_c^1(t) + \frac{I(t_2)-I(t_1)}{I(t)} x_c^2(t) + \frac{I(t) -I(t_{2})}{I(t)} x_c^3(t), & \quad {\rm if\ }t \ge t_{2}.
\end{dcases}
\end{equation}
Pulsatility is now incorporated in the model by explicitly superimposing of the two-stage dynamics of each pulse. Figures~\ref{fig:xcom_1p}--\ref{fig:xcom_3p} show the corrected model predictions. It can been seen that the oscillatory trend of the first jet-like regime is accurately predicted by leveraging the particle entrainment coefficient obtained from of each pulse instead of from the entire particle cloud. Note that this modified model can be readily applied to different coughing profiles or extended to speech patterns (which are essentially a continuous train of expulsions~\cite{abkarian2020speech,tan2021experimental}) to investigate other effects on particle penetration.

\section{Conclusion}
In this work, direct numerical simulations of particle-laden turbulent pulsatile jets were conducted to assess the role of pulsatility on particle dynamics. Realistic turbulence was provided at the jet orifice using data obtained from an auxiliary simulation of a turbulent pipe flow. The flow rate of the incoming turbulence was modulated in time according to a damped sine wave that provides control over the number of pulses, their duration, and peak amplitude. Particles were injected in the flow with diameters sampled from a lognormal distribution informed by experimental measurements from the literature.

Vortex structures were analyzed for single-, two-, and three-pulse jets. Qualitative comparison of $Q$-criterion revealed that the two-pulse and three-pulse cases exhibit multiple vortex ring structures with high vorticity regions persisting for longer times near the orifice. Entrainment coefficients were found to be larger for the multi-pulse cases compared to the single-pulse case due to the vortical structures generated by subsequent pulses, with their largest magnitude in concert with the pulses.

Particle dispersion and penetration were found to be hindered by increased pulsatility. However, particles emanating from later pulses traveled further downstream with increased pulsatility due to acceleration by vortex structures. The observed increase in penetration by later pulses may prove to be significant when determining distances at which an infectious person can be harmful to others, especially later expulsions have been found to contain higher viral concentrations.

The evolution of the particle cloud penetration was then compared to an existing single-pulse model by~\citet{bourouiba2014violent}. While the penetration for all three cases are well predicted by the puff-like regime ($x_c \sim t^{1/4}$) at late time, they deviate from the jet-like regime ($x_c \sim t^{1/2}$) at early time and instead exhibit oscillations for the pulsatile cases. A modified model was therefore proposed to account for pulsatility by leveraging the particle entrainment coefficient of each pulse and has been shown to accurately predict the oscillatory trend of the early-stage penetration.

\section*{Acknowledgements}
The computing resources and assistance provided by the staff of the Advanced Research Computing at the University of Michigan, Ann Arbor are greatly appreciated. We would also like to acknowledge the National Science Foundation for partial support from awards CBET 2035488 and 2035489.

\section*{Data Availability}
The data that support the findings of this study are available from the corresponding author upon reasonable request.
\appendix

\section{Turbulent inflow generation}\label{sec:pipe} 
To accurately model an inlet condition resembling the expiratory turbulent flow exiting from a human mouth, a direct numerical simulation (DNS) of single-phase flow traversing through a cylindrical pipe is performed. The pipe diameter $D = 0.02\,{\rm m}$ is representative of typical mouth opening. The fluid-phase equations are discretized on a Cartesian mesh, and a conservative immersed boundary (IB) method is employed to model the cylindrical pipe geometry without requiring a body-fitted mesh. The method is based on a cut-cell formulation that requires rescaling of the convective and viscous fluxes in these cells, and provides discrete conservation of mass and momentum~\cite{meyer2010conservative,pepiot2010direct}. 

\begin{figure}[!h]
\captionsetup[subfigure]{labelformat=simple}
\centering
{\includegraphics[width=0.485\textwidth]{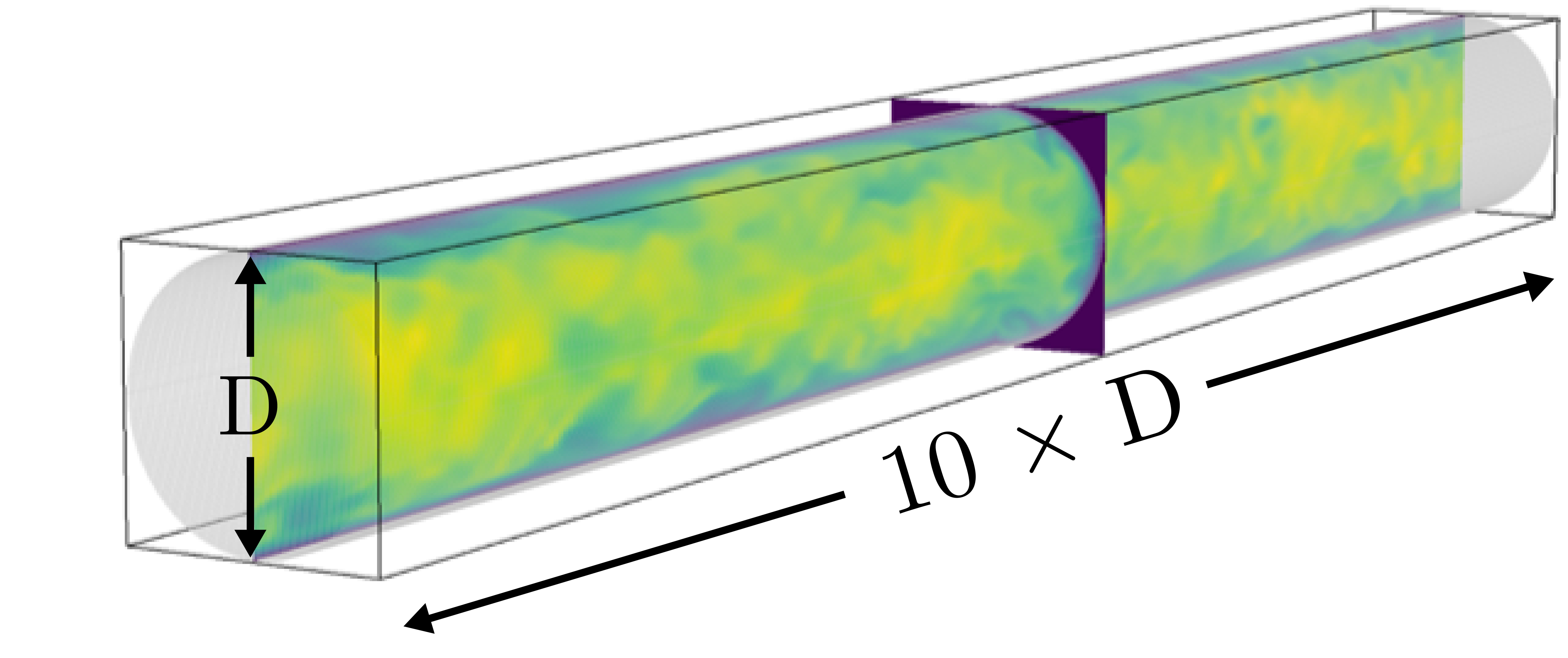}}
\caption{Instantaneous snapshot of the auxiliary DNS used to provide fully-developed turbulence at the jet inlet. Color depicts the fluid velocity magnitude.}
\label{fig:pipe_vmag}
\end{figure}

\begin{figure}[!h]
\captionsetup[subfigure]{labelformat=simple}
\centering
{\includegraphics[width=0.48\textwidth]{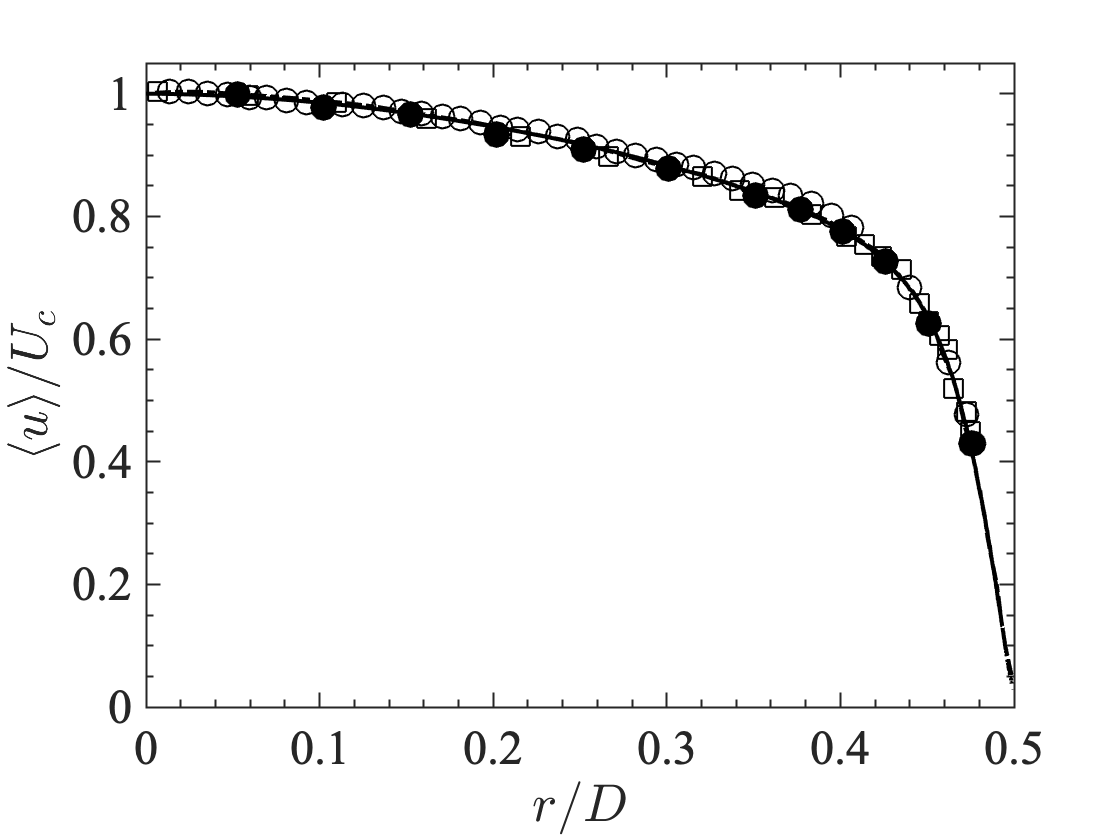}}
{\includegraphics[width=0.485\textwidth]{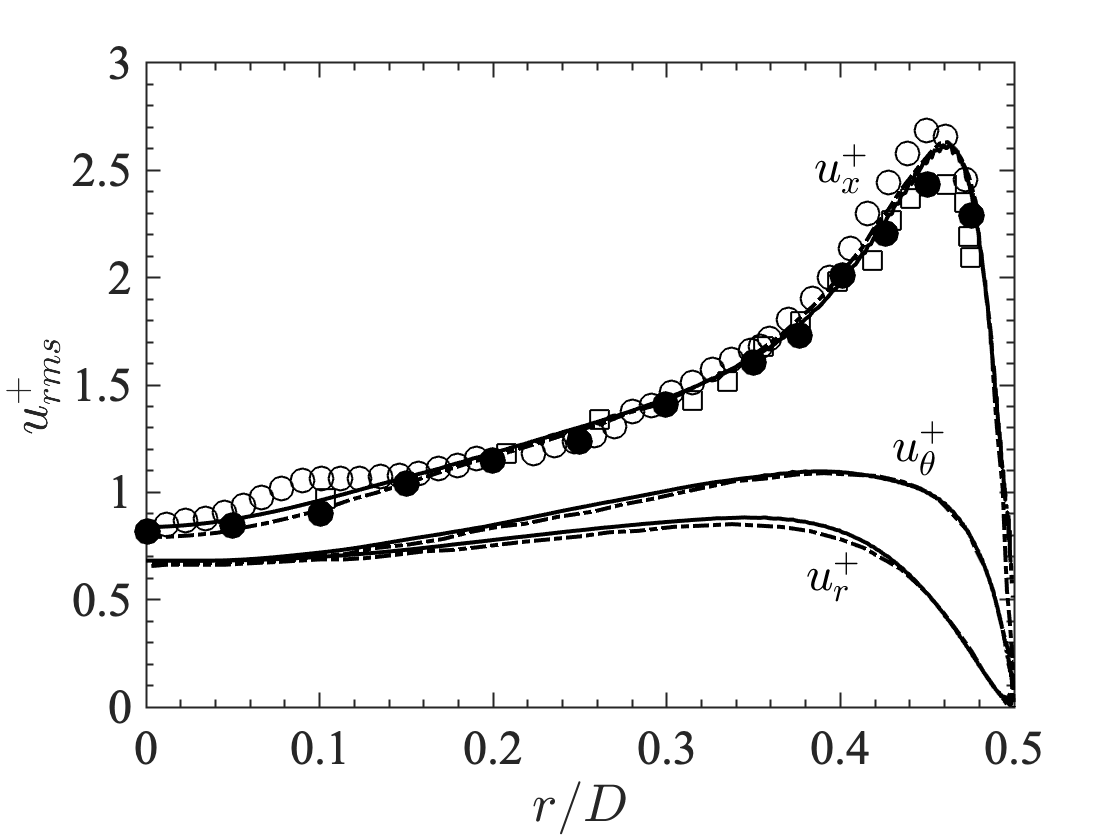}}
\caption{Mean fluid velocity profiles in fully developed turbulent pipe flow.  (a) Mean streamwise velocity normalized by the centerline value $U_c$ and (b) normalized root-mean-square velocity.  Current study ($-$), DNS ($- \cdot$) and experiments (PIV: $\Circle$, LDA: $\CIRCLE$, HWA: $\square$) by~\citet{eggels1994fully}.}
\label{fig:pipe_stats}
\end{figure}

We consider a domain of size $10D\times D\times D$, discretized using $326\times256\times256$ grid points (see Fig.~\ref{fig:pipe_vmag}). The grid spacing is chosen such that $\Delta y^{+}=\Delta z^{+}  = 1.25$ and $\Delta x^{+} = 9.8$ to satisfy the resolution criteria of DNS for pipe flows~\cite{picano2009spatial,marchioli2003direct}, where $(\cdot)^+=(\cdot) u_\tau/\nu$ denotes frictional wall units with $u_\tau$ the friction velocity. Periodic boundary conditions are applied in the streamwise direction. A uniform source term resembling a mean pressure gradient is added to the right-hand side of Eq.~\eqref{velocity} and adjusted dynamically to maintain the desired flow rate. The flow is initialized with a bulk velocity $U_0 = 4.0$ m/s with $10 \%$ sinusoidal fluctuations to accelerate the transient process. A statistical stationary state is reached after $240\,D/U_0$.  A comparison of the velocity statistics against DNS experimental data from the literature~\citep{eggels1994fully} is provided in Fig.~\ref{fig:pipe_stats}.


The velocity field at steady state is saved and used to prescribe the boundary condition in the main simulation. At each simulation timestep, the velocity in the $yz$-plane is interpolated from the fine-scale auxiliary simulation onto the coarser domain boundary of the main simulation. The fluid velocity is then rescaled to achieve the desired time-dependent flow rate according to Eq.~\eqref{damped_pulse}. The number of particles injected into the main simulation is adjusted each time step to obtain the same mass flow rate as the fluid. The velocity assigned to each particle is equal to the fluid velocity interpolated to its location. This assumes that particles expelled from the orifice have zero interphase slip velocity, and thus zero initial drag.

\bibliographystyle{apsrev4-1} 
\bibliography{ref}

\end{document}